\documentclass[preprint,journal]{vgtc}            

\onlineid{1152}

\ieeedoi{10.1109/TVCG.2023.3327149}

\newcommand{\eg}{e.g.}

\vgtccategory{Research}

\vgtcpapertype{algorithm/technique}

\title{Reducing Ambiguities in Line-based Density Plots\\by Image-space Colorization}

\author{%
  Yumeng Xue, Patrick Paetzold, Rebecca Kehlbeck, Bin Chen, Kin Chung Kwan, Yunhai Wang, and Oliver Deussen
}

\authorfooter{
  \item
    Y. Xue is with University of Konstanz, Germany and Shandong University, China.
    E-mail: {yumeng.xue}@uni-konstanz.de.
  \item
  	P. Paetzold, R. Kehlbeck, B. Chen, and O. Deussen are with University of Konstanz, Germany.
  	E-mail: {firstname.lastname}@uni-konstanz.de.
  \item
    K.C. Kwan is with California State University Sacramento, United States.
    E-mail: kwan@csus.edu.
  \item
  	Y. Wang is with Shandong University, China.
  	E-mail: cloudseawang@gmail.com.
   \item
   O. Deussen and Y. Wang are joint corresponding authors.
}

\abstract{%
 Line-based density plots are used to reduce visual clutter in line charts with a multitude of individual lines. However, these traditional density plots are often perceived ambiguously, which obstructs the user's identification of underlying trends in complex datasets. Thus, we propose a novel image space coloring method for line-based density plots that enhances their interpretability. Our method employs color not only to visually communicate data density but also to highlight similar regions in the plot, allowing users to identify and distinguish trends easily. We achieve this by performing hierarchical clustering based on the lines passing through each region and mapping the identified clusters to the hue circle using circular MDS. Additionally, we propose a heuristic approach to assign each line to the most probable cluster, enabling users to analyze density and individual lines. We motivate our method by conducting a small-scale user study, demonstrating the effectiveness of our method using synthetic and real-world datasets, and providing an interactive online tool for generating colored line-based density plots.
}


\teaser{
  \centering
     \begin{subfigure}[b]{0.32\textwidth}
         \centering
         \includegraphics[width=\textwidth]{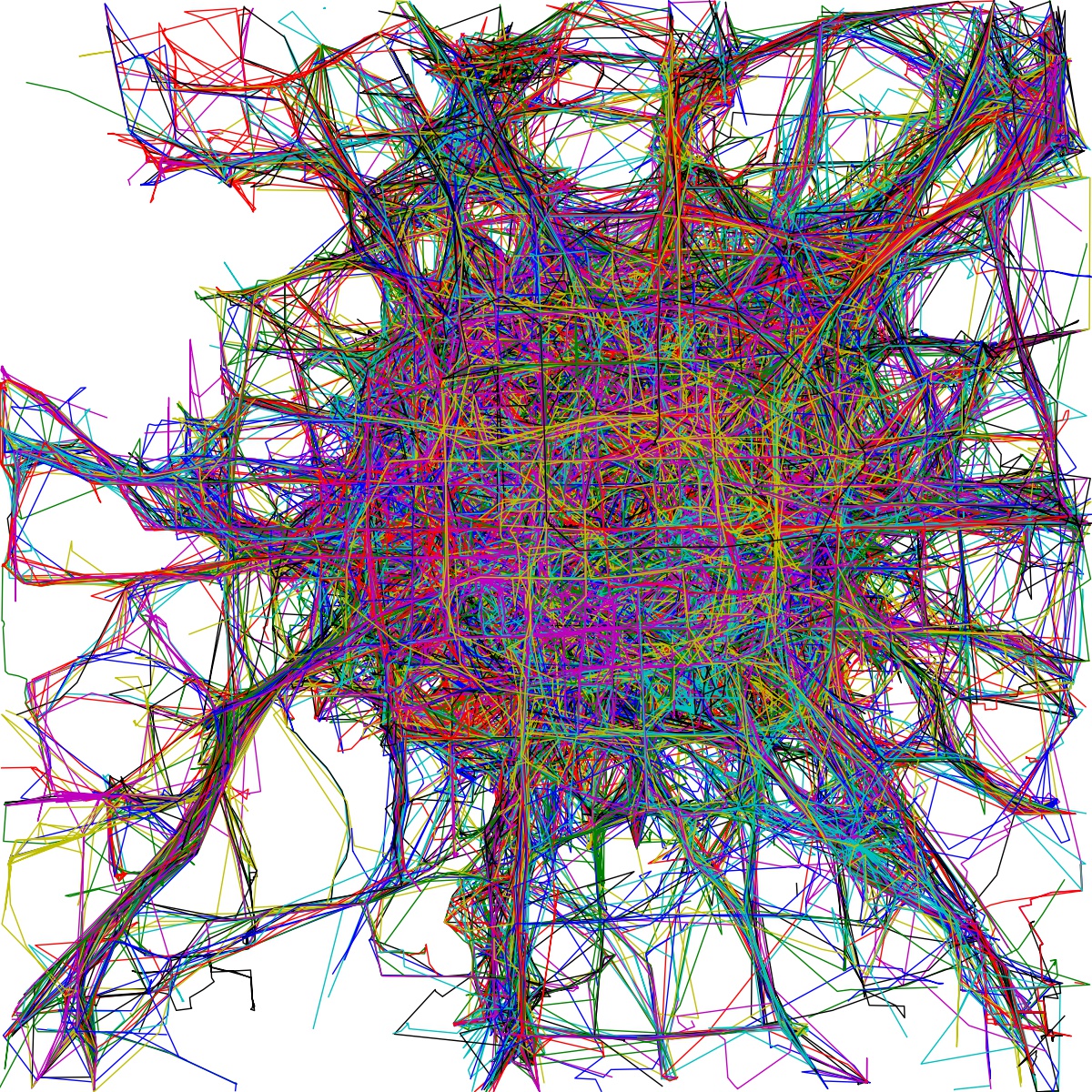}
         \caption{Line-based visualization}
         \label{fig:taxi_raw}
     \end{subfigure}
     \hfill
     \begin{subfigure}[b]{0.32\textwidth}
         \centering
         \includegraphics[width=\textwidth]{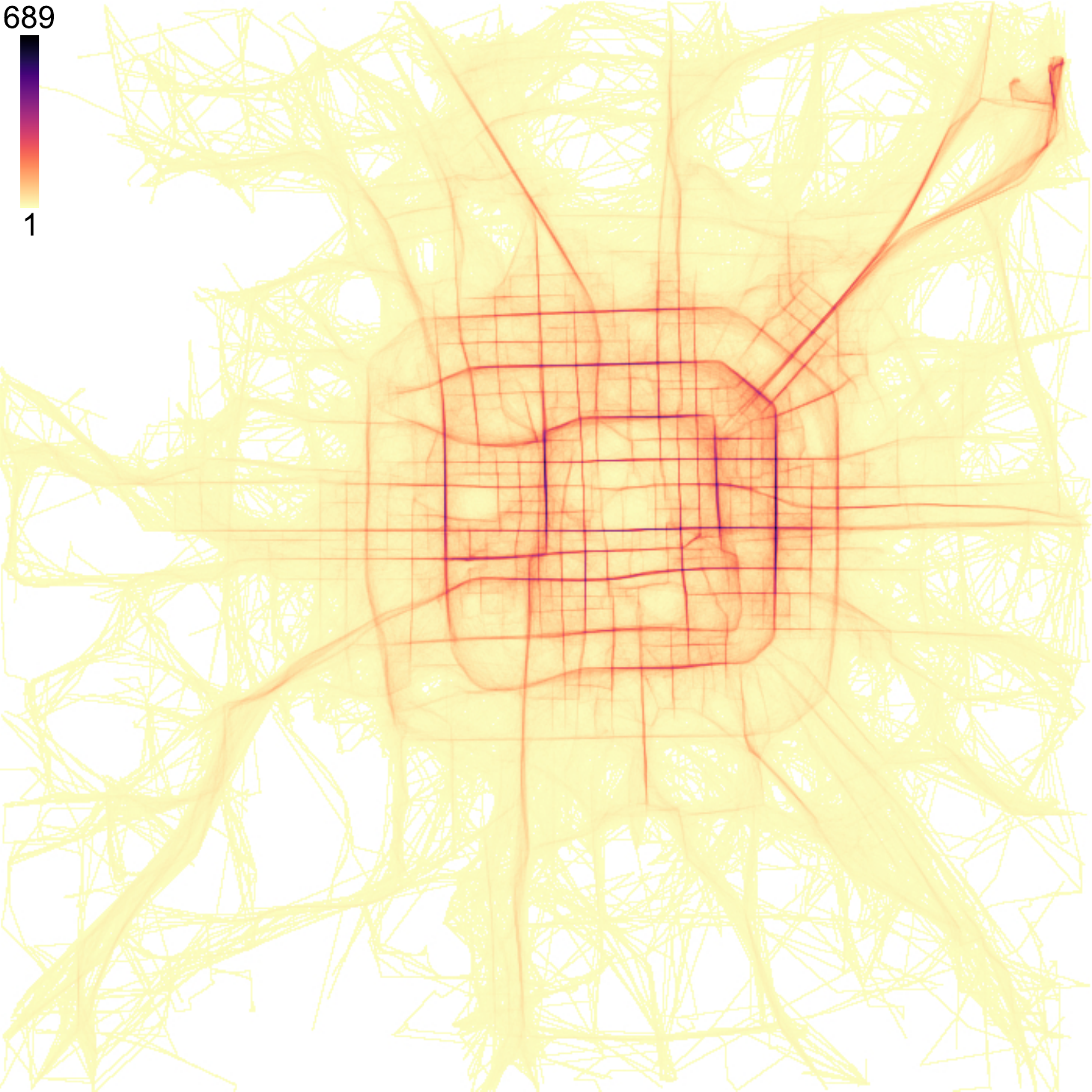}
         \caption{Line-based density plot}
         \label{fig:taxi_density}
     \end{subfigure}
     \hfill
     \begin{subfigure}[b]{0.32\textwidth}
         \centering
         \includegraphics[width=\textwidth]{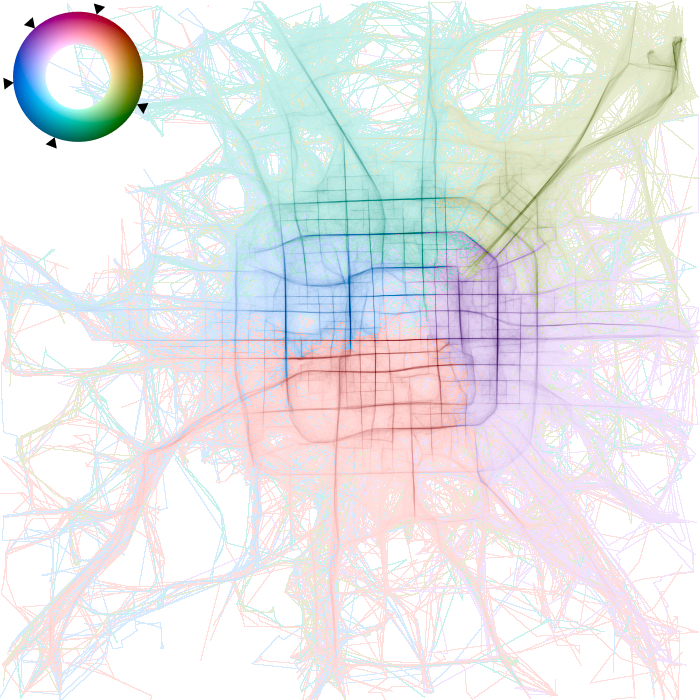}
         \caption{Ours: Colored line-based density plot}
         \label{fig:taxi_colored}
     \end{subfigure}
    \caption{Trajectories of taxi rides in Beijing: (a) A line-based visualization of the trajectories is cluttered and convoluted; (b) Using a density plot eliminates clutter, but the continuation of the trends is ambiguous, e.g., it looks like taxis follow the prominent, circular route around the city center; (c) Pixel-based colorization reveals clusters, we see that taxis mostly stay in one part of the city.}
    \label{fig:teaser}
}

\renewcommand{\manuscriptnotetxt}{}

\graphicspath{{figs/}{figures/}{pictures/}{images/}{./}} 

\usepackage{tabu}                      
\usepackage{booktabs}                  
\usepackage{lipsum}                    
\usepackage{mwe}                       

\usepackage{mathptmx}                  

\usepackage{amsmath}
\usepackage{amsfonts}
\usepackage{import}
\usepackage{color}

\usepackage{graphicx}

\newif\ifshowcomments
\showcommentsfalse

\definecolor{blue(ncs)}{rgb}{0.0, 0.53, 0.74}

\ifshowcomments
\newcommand{\todo}[1]{{\color{red}{[TODO: #1]}}}
\newcommand{\TODO}[1]{{\color{red}{[TODO: #1]}}}
\newcommand\odc[1]{{\color{cyan}{\tiny\textbf{\,[OD]\,}}{#1}}}
\newcommand\odr[2]{{\color{cyan}\sout{#1} {#2}}}
\newcommand{\revised}[1]{{\color{red}#1}}
\newcommand{\rk}[1]{{\color{blue}{#1}}}
\newcommand{\pp}[1]{{\color{orange} [PP: #1]}}
\newcommand{\ym}[1]{{\color{olive} [#1]}}
\else
\newcommand{\todo}[1]{}
\newcommand{\TODO}[1]{}
\newcommand{\revised}[1]{#1}
\newcommand{\odc}[1]{}
\newcommand{\odr}[1]{}
\newcommand{\rk}[1]{}
\newcommand{\pp}[1]{}
\newcommand{\ym}[1]{}
\fi

\begin{document}


\firstsection{Introduction}

\maketitle
Line-based plots are a popular way to visualize time series data~\cite{fulcher2013highly} and trajectory data~\cite{chen2015survey}.
These plots are widely used across domains, such as finance, healthcare, and navigation, to highlight evolving trends in complex datasets. 
To emphasize patterns and gain insights, two effective techniques are line coloring and density plots.
By using line coloring, different lines in a plot can be assigned distinct colors, making it easier to distinguish between different groups or categories. This technique is particularly useful when dealing with multiple time series or trajectories that are plotted together. Alternatively, density plots show the distribution of data at each point in time or space. These plots use color or shading to indicate areas of high or low data density, providing a more detailed picture of the underlying patterns in the data.

Line coloring assigns a color to each line. However, even in small datasets, line intersections can cause overplotting, where lines are drawn on top of each other, making it difficult to interpret the data. This problem becomes more pronounced in complex datasets with a larger number of lines, resulting in visual clutter that can obscure underlying trends~\cite{javed2010graphical}, see Fig.~\ref{fig:taxi_raw}. Mainly, this has been addressed by replacing individual lines through visual abstractions~\cite{ferstl2015streamline, Whitaker2013ContourBoxplots, mirzargar2014curve}, which use abstract glyphs to display statistical information about the lines.

Using density plots~\cite{moritz2018visualizing, lampe2011curve} shifts the focus from visualizing individual data items to aggregating data attributes in image space, communicating the structure of the data better. 
By using a continuous color map to visually communicate the density at a given position, density plots can effectively reduce visual clutter. However, the individual line information is lost. Such plots do not reveal which and how lines interact at a specific position. 
To alleviate this issue, interaction techniques have to be used (\eg, timebox~\cite{hochheiser2004dynamic}) with representative lines (such as kd-box~\cite{zhao2021kd}) being used to explore patterns of interest.

In general, due to the continuous nature of time series and trajectory data, line-based density plots are more challenging to interpret than, e.g., point-based density plots. 
Visual ambiguity within the perceived patterns often leads to false conclusions, as the Gestalt principles of good continuation and similarity strongly influence the perception of interacting lines. 
Similarly, the Principle of Unambiguous Data Depiction introduced by Kindlmann and Scheidegger~\cite{Kindlmann2014AlgebraicProcessVisualization} describes visualizations that fail to be unambiguous as having so-called ``confusers''. This means different data inputs can result in visualizations that are not distinguishable. 
\revised{For example, in Fig.~\ref{fig:taxi_density}, it looks like most taxis drive on the circular route around the city center.}  
The question is if this observed trend exists in the data or if other combinations of patterns could result in a similar visualization.
Both line coloring and traditional density plots offer no solution to help users to identify the existence, continuation, and intertwinement of trends.
Clustering the data using our method shows that there is no main circular trend around the city center; rather, taxis mostly drive within local segments of the city.

Our idea is to use an image-space coloring method to analyze line-based density plots.
In comparison to traditional density plots, we use color not just as an indicator for density but also to visually highlight similar regions in the plot. 
Each region is characterized by the lines passing through it. As a result, regions with similar lines are colored similarly, allowing users to identify and distinguish trends. 
To obtain the colors, we perform hierarchical clustering based on the lines passing through regions and subsequently map the identified cluster to the hue circle using circular MDS.
\revised{To provide additional line-based analysis, once} the bins in the density plot are clustered, we suggest a heuristic approach to allocate each line to the most probable cluster. This enables the user to explore not only the density plot but also individual lines.
We demonstrate the capabilities of our method using selected synthetic and real-world datasets. Our main contributions of this paper are:
\begin{itemize}
    \setlength\itemsep{0.25em}
    \item A novel image space coloring scheme to enrich line-based density plots with similarity information;
    \item A heuristic method to assign lines to their most likely cluster;
    \item A motivational user study to demonstrate the ambiguous perception of line-based density plots; and
    \item An interactive online tool to generate colored density plots.
\end{itemize}

In the subsequent sections, we illustrate ambiguities in the perception of trends in line-based density plots using various examples, propose our novel image-based coloring method, and perform a user study to identify ambiguities in line-based density plots.

\section{Related Work}
\label{sec:related}
In our discussion of related work, we focus on density plot generation, \revised{visual ambiguities} and illusions, line-based analysis, and color mapping techniques.
\subsection{Density-based Visualizations}
Density plots are a popular method to generate uncluttered scatter plots~\cite{carr1987scatterplot}.
There are several techniques to visualize density, including opacity blending~\cite{matejka2015dynamic}, kernel density estimation (KDE)~\cite{silverman1986density, feng2010matching}, and binning followed by summing for color mapping~\cite{wickham2013bin}.
To reduce visual clutter, density-based adaptations of various types of visualizations for different data types have been proposed.
For example, Artero {\textit{et al.}}~\cite{artero2004uncovering} used density-based filtering for parallel coordinates, while Zinsmaier {\textit{et al.}}~\cite{zinsmaier2012interactive} proposed an interactive rendering method for large-scale graphs using KDE-based node aggregation.
Scheepens {\textit{et al.}}~\cite{scheepens2011interactive} used variable KDE kernel radii for user-customizable trajectory exploration with density maps and extended their technique to combine density fields of multiple attributes in a single visualization~\cite{scheepens2011composite}.
Wickham~\cite{wickham2013bin} employed binning, summarizing, and smoothing to abstract large datasets and emphasize patterns, while Jerding and Stasko~\cite{jerding1998information} introduced a reduced representation of line charts by using gray-scale values based on the level of overlap.
Lampe {\textit{et al.}}\cite{lampe2011curve} extended KDE to curves and called their method curve density estimates (CDE).
Recently, Moritz and Fisher~\cite{moritz2018visualizing} proposed DenseLines, a discretized CDE variation that allows parallel computing on the GPU.
However, all these works focus on calculating pixel density and do not consider the relationships between pixels.
\subsection{\revised{Visual Ambiguities} and Illusions}
Much research has been done to tackle \revised{visual ambiguities} or illusions that lead humans to draw wrong conclusions.
Feng {\textit{et al.}}~\cite{feng2010matching} visually encoded uncertainty present in the data in scatter plots and parallel coordinates to prevent users from drawing false conclusions about the data.
Pomerenke {\textit{et al.}}~\cite{pomerenke2019slope} explored the relationship between slope and the perceived prominence of lines in ghost clusters and proposed a slope-dependent density correction method to reduce visual errors.
For scatter plots, Liu {\textit{et al.}}~\cite{liu2021data} refine the orientation of marks to guide the users in estimating trends to prevent imprecise estimations biased by human vision.
Hong {\textit{et al.}}~\cite{hong2021weighted} studied how the size and lightness in scatterplots affect the perceived mean.
Some other methods use additional visual coding to reduce \revised{visual ambiguities} and illusions.
Novotn\'y and Hauser~\cite{novotny2006outlier} proposed an outlier preserving method for parallel coordinates. They extract outliers by binning in the adjacent dimensions and detect outlier bins.
Trautner {\textit{et al.}}~\cite{trautner2020sunspot} highlight outliers by overlaying additional visual coding on the scatter-based density map. Unfortunately, there is no current work that focuses on \revised{visual ambiguities} and illusions in line-based density plots.
\subsection{Line-based Analysis}
Line-based data analysis methods use lines as fundamental units of data and enable various analyses, including clustering, abstraction, and interactive exploration. Line clustering \revised{uses} different metrics~\cite{zhang2006comparison} to measure the distance between 2D or 3D lines, such as the Euclidean distance~\cite{chen2007similarity, rossl2011streamline}, curvature or torsion~\cite{mcloughlin2012similarity, yu2011hierarchical}, and user-specified streamline predicates~\cite{salzbrunn2006streamline}. Dynamic time warping (DTW)~\cite{muller2007dynamic} is another widely used metric to measure similarity between time series data. The clustering of lines can be achieved by applying these similarity metrics.
To obtain cluster centers for clustering time series data, Petitjean {\textit{et al.}}~\cite{PETITJEAN2011678} proposed a global averaging method based on DTW, while Gaffney and Smyth~\cite{gaffney2004joint} presented a framework for clustering curves based on probabilistic curve alignment models.
Some methods use Euclidean space vectors to represent the line curves and then apply Euclidean distance-based clustering methods~\cite{chen2011illustrative, rossl2011streamline}. 

However, even after obtaining clustering information, further analysis remains a challenge because lines can still be cluttered, and clustering methods are sensitive to noise.
Uncertainty visualizations therefore use glyphs to visualize statistical information and enable overviews.
Mirzargar {\textit{et al.}}~\cite{mirzargar2014curve} generalized the curve boxplot for ensembles of curves by introducing a functional band depth.
Ferstl {\textit{et al.}}~\cite{ferstl2015streamline} applied clustering to construct curve-boxplot-like abstractions of multiple line bundles.
Palmas {\textit{et al.}}~\cite{palmas6787137} abstracted clusters in parallel coordinates through edge bundling, facilitating further interaction with clusters.
These visualization methods are effective only for regularly distributed lines and are not suitable for noisy datasets.

\revised{In addition to the previously introduced clustering approaches that either visually cluster lines by color or abstract them, edge bundling techniques \cite{holten2006hierarchical, holten2009force, zwan2016cubu} alter the course of individual lines to combine them into bundles to reduce visual clutter. To incorporate geographical restrictions like roads, specially tailored edge bundling techniques~\cite{thony2015vector, zeng2019route} were proposed. Recently Wallinger {\textit{et al.}}~\cite{wallinger2021edge, wallinger2023faster} proposed an edge-path bundling method that reduced the ambiguities introduced by edge bundling.}
\revised{
Our method is focused on reducing the ambiguity while not altering the original data, as edge bundling does.}

Interaction methods are also important for analyzing lines.
For example, QuerySketch~\cite{wattenberg2001sketching} allows users to sketch freely and query the data lines that match the shape of the sketched line. Hurter {\textit{et al.}}~\cite{hurter5290707} introduced interactive paradigms to extract trajectories of interest from large-scale airline trajectory data. Hochheiser and Shneiderman~\cite{hochheiser2004dynamic} presented Timebox to query data that passes through a box, representing a range of positions within a certain time.
Recently, Zhao {\textit{et al.}}~\cite{zhao2021kd} proposed KD-Box by using a KD-tree to speed up the timebox query and introduce representative lines to assist analysts in exploring the local pattern details of interest.
However, these methods require significant user interaction and learning costs, and the small number of representative lines may not fully reflect the overall composition of patterns.
\begin{figure*}[ht]
     \centering
     \begin{subfigure}[b]{0.26\textwidth}
         \centering
         \includegraphics[width=\textwidth]{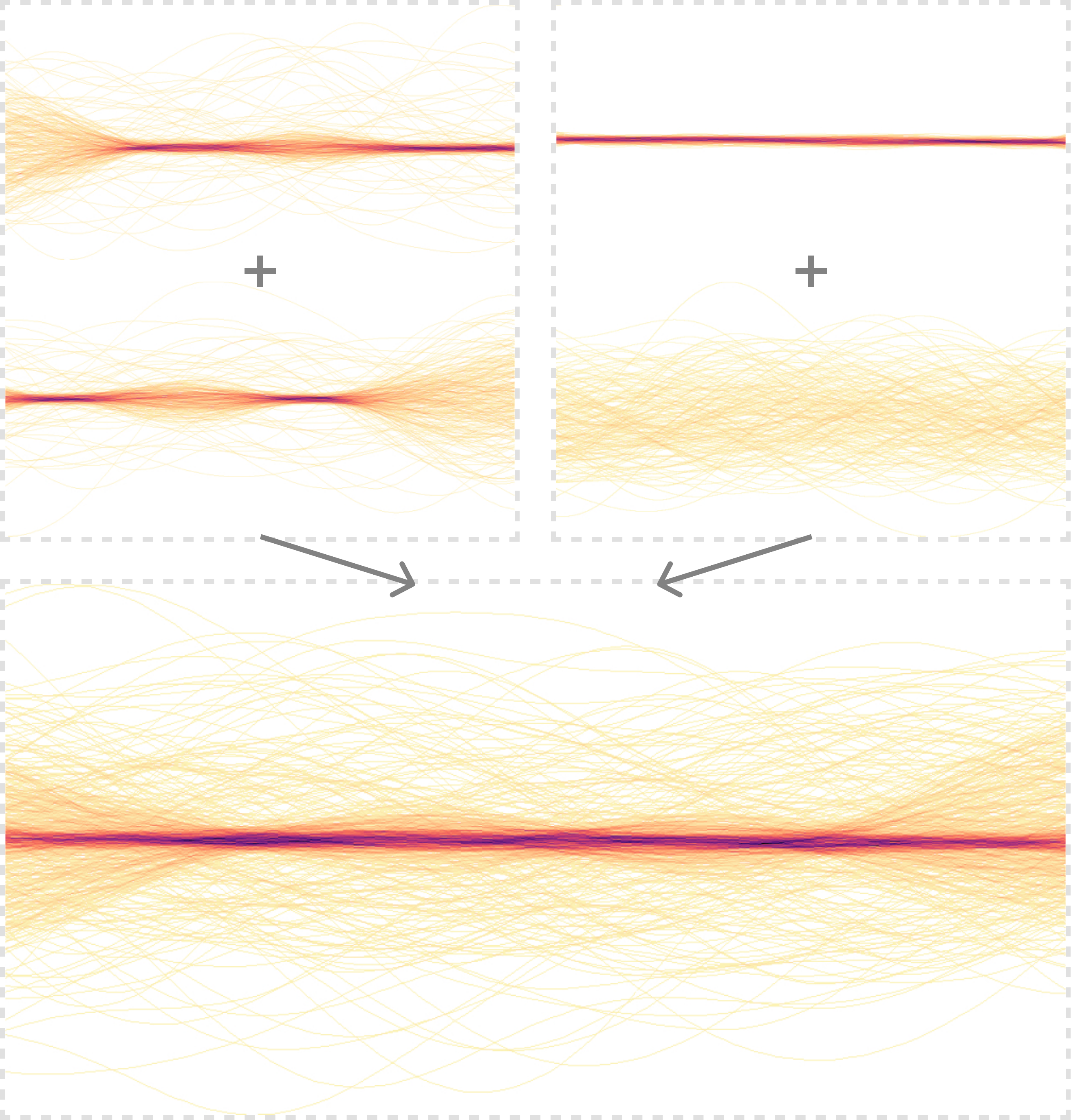}
         \caption{Illusory patterns}
         \label{fig:ambiguous_patterns_illusionary}
     \end{subfigure}
     \hfill
     \begin{subfigure}[b]{0.26\textwidth}
         \centering
         \includegraphics[width=\textwidth]{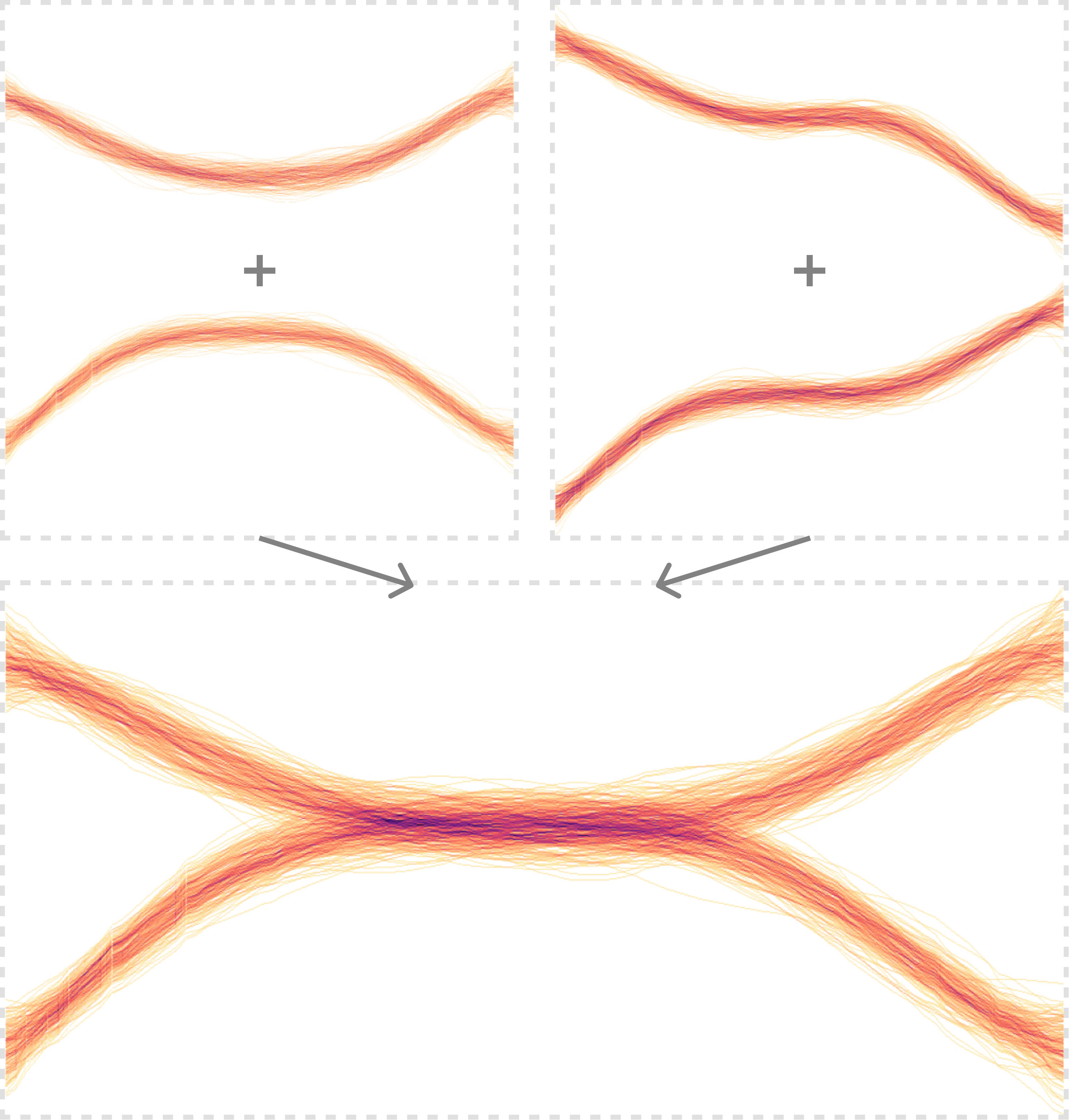}
         \caption{Ambiguous continuation}
         \label{fig:ambiguous_patterns_continuation}
     \end{subfigure}
     \hfill
     \begin{subfigure}[b]{0.26\textwidth}
         \centering
         \includegraphics[width=\textwidth]{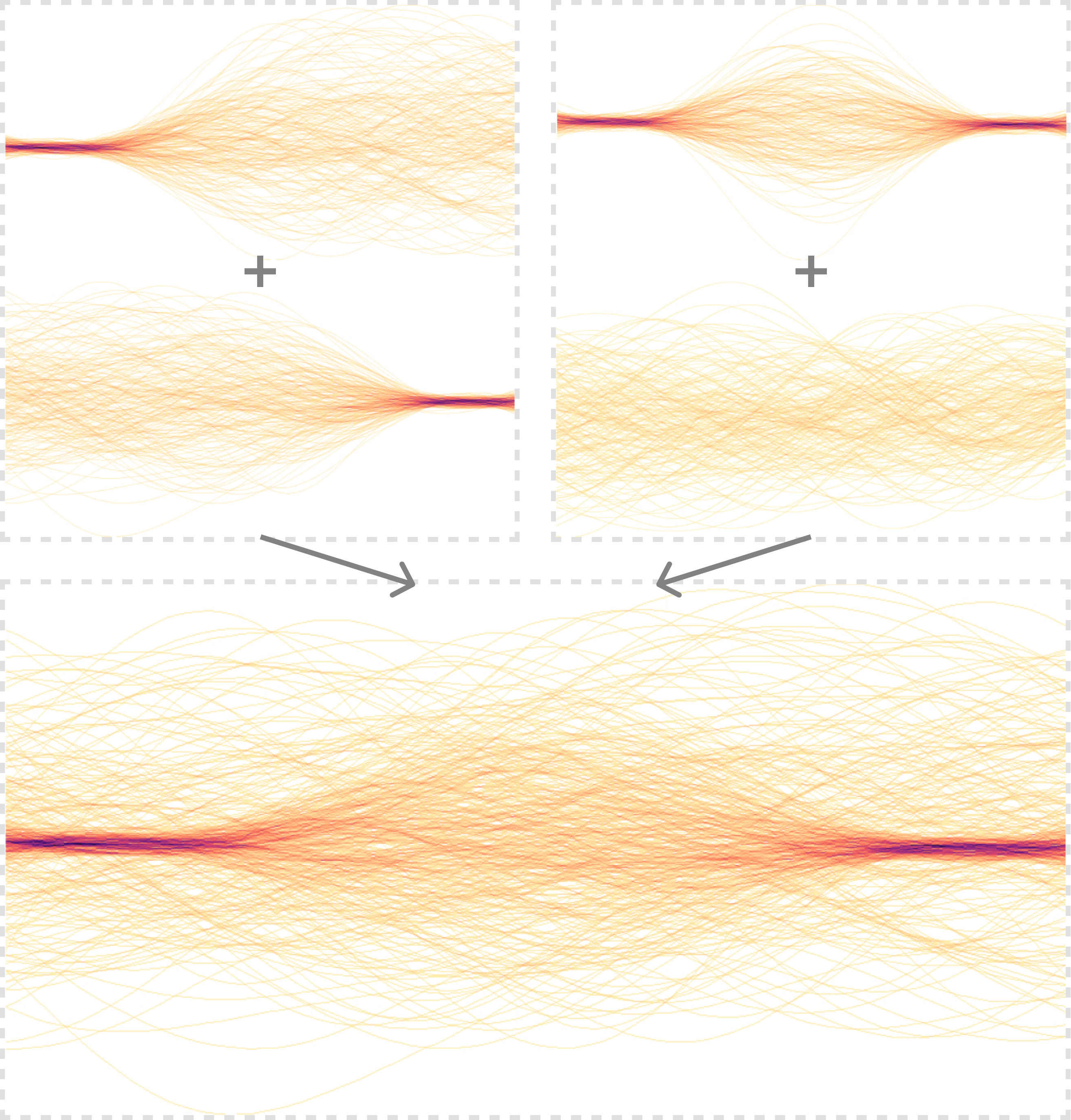}
         \caption{Disconnected clusters}
         \label{fig:ambiguous_patterns_disconneted}
     \end{subfigure}
    \vspace{-3mm}
    \caption{Three examples of ambiguities in line-based density plots. Each density plot in the lower row could be a superposition of one of the two trends in the upper row. (a) The density could come from two trends with varying densities or from a clear trend and added noise. (b) The pattern could be two touching trends or two crossing ones. (c) The trend could be a single, continuous trend, or a combination of two trends that diverge.}
    \vspace{-4mm}
    \label{fig:ambiguous_patterns}
\end{figure*}
\subsection{Color Mapping}
Color maps are an important aspect of data visualization, as they have a significant impact on how effectively the viewer perceives and understands the underlying data. Zhou and Hansen~\cite{zhou2015survey} conducted a comprehensive survey of color map generation methods and classified them into four categories: procedural-based~\cite{robertson1986generation, trumbo1981theory, pham1990spline, moreland2009diverging}, user-study-based~\cite{healey1996choosing, rogowitz2001blair, kindlmann2002face}, rule-based~\cite{bergman1995rule, rheingans2000task, tominski2008task}, and data-driven~\cite{thompson2013provably, wang2012importance, tennekes2014tree, lee2012perceptually}.
In the context of density plots, sequential color maps are typically used to display the gradual changes in density values. The Brewer color palettes, established by Brewer {\textit{et al.}}~\cite{brewer1994color, brewer1994guidelines, brewer1999color}, provide a set of commonly used sequential, diverging, and qualitative color maps and guidelines for selecting appropriate colors for different types of visualizations.

Color maps may introduce visual artifacts or lack perceptual consistency.
Perceptually consistent color maps are designed to match the human perception of color and are constructed using color spaces such as HCL~\cite{hornik2003colour}.
Zeileis {\textit{et al.}}~\cite{zeileis2009escaping} proposed a perceptually consistent single-hue color map for  density visualizations.
Lu {\textit{et al.}}~\cite{lu2020palettailor} proposed Palettailor to optimize categorical colorization.
It can be used to obtain perceptually consistent colors while maintaining the discriminability of positionally close categories.
\revised{However, coloring methods for categorical data (e.g., multiclass scatterplots) are not applicable to our problem, as the colors they assign do not carry density information.
Our goal is twofold: we want to visualize a cluster membership while retaining density information.
Therefore, multiple single-hue colormaps based on HCL color space are well suited for this problem.
They convey clustering information using hues while maintaining visual consistency for the same density.
In addition, the hue harmonic method proposed by Cohen-Or~{\textit{et al.}}~\cite{cohen2006color} can be used to further optimize the hue selection.
}

\section{Density Plots of Lines}
We follow the process proposed by Moritz and Fisher~\cite{moritz2018visualizing} to create the line-based density plot. Firstly, the visual area is divided into bins (in the smallest case, a bin is one pixel). For each bin, the number of entities --- points in scatter plots and lines in line charts --- are counted, and the density is mapped to a sequential color palette.
This is the application of Wickham's bin-summarize-smooth~\cite{wickham2013bin} paradigm to line data.
In the subsequent paragraphs, we provide a detailed overview of sources of ambiguity in the interpretation of line-based density plots.
In Section \ref{sec:user-study}, we use the three defined sources of ambiguities to show in a small-scale motivational user study how participants perceive ambiguous line-based density plots.

\subsection{Time series and Trajectory Definition}
Mirzargar {\textit{et al.}}~\cite{mirzargar2014curve} define trajectories mathematically as parametric curves mapping from a domain $\mathbf{D}$ to a potentially higher dimensional co-domain $\mathbf{R}$.
As we aim to generate two-dimensional density plots, we restrict our consideration of trajectories to one and two-dimensional co-domains. For one-dimensional co-domains, the trajectory is a time series representation. Two-dimensional co-domains can represent, for example, geospatial data, like traffic data or human movements, or more abstract spaces, like eye-tracking trajectories.

\begin{figure*}[ht!]
    \centering
     \includegraphics[width=\linewidth]{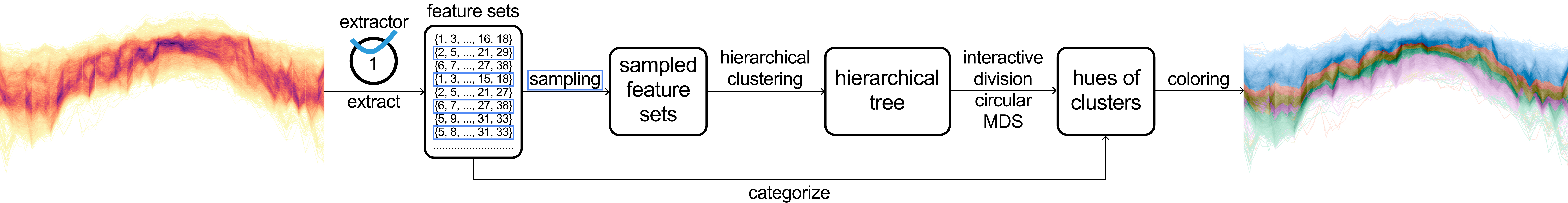}
    \vspace{-3mm}
    \caption{Processing pipeline: per region (bin) sets of features (line IDs) are obtained by checking which lines pass through a bin. By sampling, we obtain a subset of the bins that allow us to cluster them efficiently.
    Hierarchical clustering is then applied to find similar bins. Using circular MDS we map cluster centers to the HCL color space to create the rendition.}
    \vspace{-4mm}
    \label{fig:pipeline}
\end{figure*}

\subsection{Ambiguities in Density Plots}
\label{sec:illusorypatterns}
While phenomena like \revised{visual ambiguities} and visual artifacts introduced by scatter plots \cite{Mayorga2013Splatterplots} and line-based plots have been studied~\cite{pomerenke2019slope}, little research has been done on density plots.
Compared to density representations of scatter plots, which are straightforward to interpret, the interplay of multiple lines in a line-based density plot might add visual ambiguities hampering their clarity.
\revised{Investigating} line-based density plots of real-world datasets, we found patterns with ambiguities and further \revised{compiled a list of three abstracted types of patterns} (Fig.~\ref{fig:ambiguous_patterns}). \revised{We focus ourselves on these three recurring patterns in our investigation and do not claim the list to be complete. In Section \ref{sec:user-study}, we show the results of our user study, underpinning that users perceive line-based density plots ambiguously.}
The upper rows of each subfigure show possibilities of individual trends that, when combined, form a similar pattern.
Even if only two trends are superimposed, it is challenging to decide which individual trends the density plot is composed of.\\
\noindent \textbf{Illusory patterns.}
The combined density plot in the lower row of Fig.~\ref{fig:ambiguous_patterns_illusionary} is visually dominated by the central horizontal line. An observer could assume this dominant region is superimposed on a less dense but noisy region surrounding it. This perceived continuation is also supported by the Gestalt principle of good continuity \cite{wertheimer1938laws}. Our example illustrates that this perception is deceptive, as the observed pattern is actually a composition of the two patterns on the top left.
This ambiguity is often present in real-world datasets, such as ship trajectories before and after passing through a strait (see Fig.~\ref{fig:usecase_Greece}).
\revised{We refer to patterns that appear to have a high-density trend only in the combined line-based density plot that is not present in the individual underlying patterns as ``illusory patterns''.}\\
\noindent \textbf{Ambiguous continuation.}
If multiple independent, well-separated trends partly overlap with a similar orientation in the overlapping region, their continuation is ambiguous. 
In the lower row of Fig.~\ref{fig:ambiguous_patterns_continuation}, it is not clear how the individual trends continue after intersecting. Both combinations on the top could create such a pattern. 
Thus, their continuation remains unclear.
This ambiguity frequently appears in data we have observed, such as time series data with crossovers and overlaps between trends~(see Fig.~\ref{fig:usecase_real_world_temerature}).\\
\noindent \textbf{Disconnected clusters.}
Dense regions in plots attract attention. If a line-based density plot contains multiple dense regions, it is ambiguous if they belong to the same trend. 
A line-based density plot with two such dense regions is shown in the lower row of Fig.~\ref{fig:ambiguous_patterns_disconneted}. Due to their visual saliency and similar density, they appear to be related.
Perceptual psychology explains this with the Gestalt principle of similarity.
However, this impression is deceptive since the pattern is actually composed of the two completely independent patterns on the left of the upper row of Fig.~\ref{fig:ambiguous_patterns_disconneted}.
Thus it is ambiguous if the combined density plot consists of two line bundles fanning out in the center of the plot or two independent trends.
It is common to see disconnected clusters in real-world datasets; an example of this can be observed in Fig.~\ref{fig:usecase_real_world_stocks}.
Even these straightforward examples show that interpreting \revised{line-based} density plots is challenging, and identifying the underlying trends is often ambiguous. Patterns in real-world data are even more complex than our examples.
We acknowledge the possibility of the existence of other types of ambiguities, which is an interesting topic for further research.
Already just the three types which we identified show that there is a need \revised{to visually better convey} trends and patterns. This is the motivation for us to propose our density plot coloring method that helps to identify individual patterns within line-based density plots.
\revised{Our method assigns similar colors to regions with similar lines passing through (Figs.~\ref{fig:illusionary_comparison_colored}, \ref{fig:ambiguous_trend_comparison_colored}, and \ref{fig:disconnected_comparison_colored}), which helps users intuitively understand the true composition of the related line patterns and their relationships.}

\section{Method}
Our goal is to reduce ambiguities in line-based density plots, identifying and highlighting similar regions.
Previously introduced data-space clustering methods cluster individual data items, like lines or scatter points. In contrast, we apply clustering in the image-space,  grouping regions based on the lines passing through them.

Generally, creating density plots involves three main steps: discretizing the canvas into 2D bins, counting the data items that touch each bin, calculating the bin's density, and optionally applying a smoothing method, e.g., a kernel method. This bin-summarize-smooth paradigm was introduced by Wickham~\cite{wickham2013bin}.
Our method assigns colors to bins in the image space based on their similarity. Similar bins are grouped into clusters with hue values assigned to them.

Our pipeline (see Fig.~\ref{fig:pipeline}) consists of four main steps: bin-based feature set extraction, reduction of the number of bins by sampling and thresholding, hierarchical clustering, and finally, the assignment of colors to bins. We implemented our method as a web-based analysis tool that offers various user-controllable parameters to analyze the data and steer the colorization.

\subsection{Feature Set Extraction}

As a first step, we discretize the plot's canvas into bins.
We see a bin as an atomic unit of the density \revised{plot}, which can contain multiple pixels, but each bin can only have one color, i.e., the number of bins can be considered as the logical resolution.
For simplicity, in the following, a bin is equivalent to a pixel. 
We define a feature set for each bin that contains the identifiers of all lines touching it.
This feature set expresses the relationship between a bin $p$ and every line $L = \left\{L_1, L_2, \dots, L_N \right\}$ in the dataset.
It encodes which lines of $L$ are spatially close to the center of $p$. For each bin, we obtain a feature set $\operatorname{S}(p)$:
\begin{equation}
    \operatorname{S}(p) = \left\{ i \mid L_i \in L \land \operatorname{distance}(L_i, p) < T \right\}.
 \end{equation}
where $\operatorname{distance}(L_i, p)$ is the closest distance between line $L_i$ and center of bin $p$. The parameter $T$ is conceptually the radius of a disc positioned at the center of $p$, defining the extraction area for that bin.
As default, we set $T=1$. 
Due to the circular distance measure, the disc created by $T$ may cover a slightly larger area than the bin.
The feature set $\operatorname{S}(p)$ then contains the identifiers of all lines running through the disc at position $p$. Increasing $T$ expands the extraction area of bins, which will result in more similar feature sets for adjacent bins.

\subsection{Bin Sampling}
\label{sec:sampling}
If the bins correspond to individual pixels, the number of feature sets $\operatorname{S}(p)$ is equal to the number of pixels in the density plot. Even a medium-resolution density plot with a resolution of $1000\times 500$ pixels contains 500,000 bins. 
In the subsequent step of our pipeline, we use agglomerative hierarchical clustering~\cite{jain1999data} to group bins based on their similarity. It has a time complexity of $O(n^2 \log(n))$, which makes it challenging to use it to cluster such a large number of bins.
One potential way to reduce the number of bins is to increase the bin size.
However, this inevitably sacrifices details and results in a loss of resolution.
Alternatively, we can either sample bins or exclude bins of low-density regions. Using sampling has the advantage that we can exploit the continuous nature of the line-based datasets that underlay the visualization - feature sets of neighboring bins are typically similar due to the spatial continuity of lines. Thus sampling does roughly retain the distribution of the feature sets.
In addition, users are typically interested in sufficiently dense regions. Low-density bins (e.g., those touching fewer than ten lines) typically contain a lot of noise and are less noticeable to the user. To filter out such bins, we provide a user-customizable minimum density threshold.

\subsection{Hierarchical Clustering}
We group the sampled bins using agglomerative hierarchical clustering.
To cluster the bins, we have to use a distance metric that is appropriate for set-type data. 
Commonly employed are the Jaccard index $\operatorname{J}(A,B) = {\frac{|A \cap B|}{|A \cup B|}}$ and Sørensen-Dice coefficient $\operatorname{DSC}(A,B)={\frac {2|A\cap B|}{|A|+|B|}}$
for sets $A$ and $B$. 
A ratio close to 1 indicates a high degree of similarity.
As densities across line plots can vary greatly, the sizes of the feature sets also differ heavily.
Because the Jaccard index and Sørensen-Dice coefficient are susceptible to variations in the set size, their values tend to be lower when comparing sets of vastly different sizes, even if one is a subset of the other.
Therefore, for line-based density plots the overlap coefficient~\cite{vijaymeena2016survey} seems to be more suitable:
\begin{equation}
    \operatorname{overlap} (A,B)={\frac {|A\cap B|}{\min(|A|,|B|)}}
\end{equation}
It is insensitive to differences in set sizes as the denominator is solely determined by the smaller set.
In a line-based density plot, the feature set of a bin in a low-density region (with dispersed lines) is typically a subset of the feature set of a bin in high-density areas. Because lines passing through a lower-density bin are likely to also pass through a higher-density bin, where the lines are more aggregated.
In this case, we expect the two sets to have a high degree of similarity. 
The overlap coefficient accurately reflects this phenomenon. \revised{A more detailed comparison is available in Section 2 of the supplementary material.}

We use average linkage hierarchical clustering~\cite{day1984efficient} \revised{as} it \revised{can be used with} arbitrary distance metrics, and the number of clusters does not have to be specified in advance. Additionally, the user can interactively divide the clusters to analyze the data further. We did not use well-established partitioning clustering methods such as PAM to identify the clusters since, here, the number of clusters has to be known in advance.

\subsection{Cluster Assignment}
\label{subsec:cluster_assignment}
In the previous step, we grouped the sampled bins into clusters based on their similarity. 
However, the bins that were not part of the sample have not been attributed to a cluster as of now.
To assign clusters to these bins, we have to calculate the similarity between clusters and bins, which is not easily possible using the above-mentioned overlap measure.
Since only the similarity between bins can be calculated, we would have to calculate the overlap coefficient similarity of an unassigned bin to all bins of each cluster, which is time-consuming.
To solve this, we instead compute a mean feature vector $\operatorname{M}(C)$ for each cluster $C$. The feature vector $\operatorname{V}(p)$ of the feature set $\operatorname{S}(p)$ of a bin is a binary vector with entries set to 1 if the identifier $i$ of the line $L_i$ is contained in $\operatorname{S}(p)$ and 0 otherwise:
\begin{equation}
\operatorname{V}(p) = (v_0^p, v_1^p, \dots, v_n^p), \quad
v_i^p = \begin{cases}1, \quad i \in \operatorname{S}(p) \\ 0, \quad \text {otherwise. }\end{cases}
\end{equation}
The mean feature vector of a cluster $C$ is defined as the element-wise sum of the binary feature vectors $\operatorname{V}(p)$ of all its bins $p$, divided by the number of bins in $C$.
\begin{equation}
\begin{aligned}
\operatorname{M}(C) &= \frac{1}{|C|}\sum_{p \in C} \operatorname{V}(p)\\
\end{aligned}
\end{equation}
To assign a cluster to each bin, we identify the most similar cluster $C$ for this bin $p$. To calculate the distance between the bin's feature vector $\operatorname{V}(p)$ and the mean feature vector $\operatorname{M}(C)$ of the cluster $C$, we use the Euclidean distance as a similarity measure. Note that when calculating the distance, we only consider the position where $v^p_i$ is 1:
\begin{equation}
    \operatorname{D}(p, C) = \sum_{i \in \operatorname{S}(p)}\left( 1 - \operatorname{M}(C)_i \right)^2
\end{equation}
This is similar to our rationale for choosing the overlap coefficient because if we consider all $v^p_i$, it would also count the lines not touching the bin, which leads to an unreasonable increase in the distance.

\subsection{Cluster Colorization}
\label{subsec:coloring}
In the next step, we assign colors to clusters so that these colors visually represent the similarities among them -- bins of similar clusters should be colored similarly, while dissimilar clusters should be colored differently. \revised{Thus, a bin's color solely depends on the cluster it is assigned to and does not, e.g., reflect the orientation of the lines passing through it.}
A bin's color should visually communicate both the bin's assignment to a cluster and the density of the bin.
Changing a bin's cluster assignment should not drastically alter its perceived density. Therefore, we use the perceptually uniform color space HCL (Hue-Chroma-Luminance)~\cite{hornik2003colour} that uses hue, chroma, and luminance as its dimensions. Since it is derived from perception science~\cite{zeileis2009escaping}, it is increasingly used for visualization. Luminance can be altered independently of chroma and hue~\cite{zeileis2009escaping}.
Different clusters with the same density should ideally be perceptually comparable to enable the user to compare densities across multiple clusters colored with different hues. Mapping density only to luminance would allow for an easy comparison of densities, but the resulting color map would not be rich in contrast. Similar to \cite{zeileis2009escaping}, we map the density of a bin simultaneously to chroma and luminance. It seems more desirable to map densities to a contrast-rich color map that matches the typical characteristics of density plot colorizations, such as darker colors for high density and light colors lower density.

Hue values can be represented as values on a hue circle in the range $[0, 2\pi]$. Therefore we need a mapping from the high dimensional mean feature vector $\operatorname{M}(C)$ of a cluster to the hue circle that preserves the similarity between feature vectors. 

Circular multidimensional scaling (MDS) is a nonlinear, non-metric dimensionality reduction method that was developed for exactly this application. It differs from the commonly used MDS, which uses the Euclidean metric. 
The stress measure in this variant of MDS~\cite{cox2008multidimensional} is defined as follows:
\begin{equation}
S = \sqrt{\frac{\sum_{i<j}\left(\delta\left(x_{i,j}\right)-d_{i,j}\right)^2}{\sum_{i<j} d_{i,j}^2}}
\end{equation}
Here, $\delta(x_{i,j})$ is the linear scaling of the distance between points $i$ and $j$ to the distance on the circle, and $d_{ij}$ is the distance between points $i$ and $j$ after dimensionality reduction to the unit circle. In contrast to the approximate distances used by TF Cox~{\textit{et al.}}~\cite{cox1991multidimensional}, the $d_{ij}$ in our case is the accurate distances along the circular arc:
\begin{equation}
    d_{ij} = 
    \begin{cases}
        \left| \theta_i - \theta_j \right|,  &  \left| \theta_i - \theta_j \right| \leq \pi \\
         2\pi - \left| \theta_i - \theta_j \right|,  &  \left| \theta_i - \theta_j \right| > \pi
    \end{cases}
\end{equation}
Therefore, for the unit circle, $\delta(x_{i,j}) \in [0,\pi]$. We use the partial derivative $\frac{\partial S}{\partial \theta_{k}}$ (details see the \revised{Section 1 of the} supplementary materials) of the stress $S$ with respect to the angle $\theta_k$ of a point $k$ for allowing gradient descent. By doing so we optimize the angle of each point and this way determine the hue value corresponding to each cluster center. 

\subsection{Cluster-based Line Filtering}
Our coloring method identifies the main trends and highlights them through colorization. While being able to analyze individual trends in isolation and identifying the lines contributing to them is an important task, traditional density plots and our colorization aggregate lines in bins and thus do not directly provide a correspondence between trends in the density plot and individual lines. Therefore, we have to assign each line to a cluster of the density plot.
To determine the cluster of a line, we focus on two aspects: grouping the bins of each cluster and summing up their weights. 
Counting the bins of each cluster that a line passes through already gives us a good intuition about which cluster may describe the line best.
However, the perceived trends mainly depend on high-density regions, so bins of these regions should be more influential.
Inspired by density-based edge clustering~\cite{lhuillier2017state} and representative line selection~\cite{zhao2021kd}, we, therefore, also consider the density of the bins a line passes through. 
Bins are expressed as:
\begin{equation}
C^i_k = \left\{ p_j \mid p_j \in C_i \land i \in \operatorname{S}(p_j) \right\}
\end{equation}
where $C^i_k$ represents the set of bins of the $k$-th cluster $C_k$ that line $i$ passes through. Then, we compute the sum of densities of the bins in the set $C^i_k$ as follows:
\begin{equation}
W^i_k = \sum\limits_{p_j \in c^i_k} D(p_j)
\end{equation}
where $D(p_j)$ is the density of bin $p_j$. Thus, we assign line $i$ to the cluster $C_k$ with the greatest summed-up weight $W^i_k$. 
This allows users to analyze the details of the lines matched to each cluster.

\begin{figure*}[p]
    \centering
     \begin{subfigure}[b]{0.32\textwidth}
         \centering
         \includegraphics[width=\textwidth]{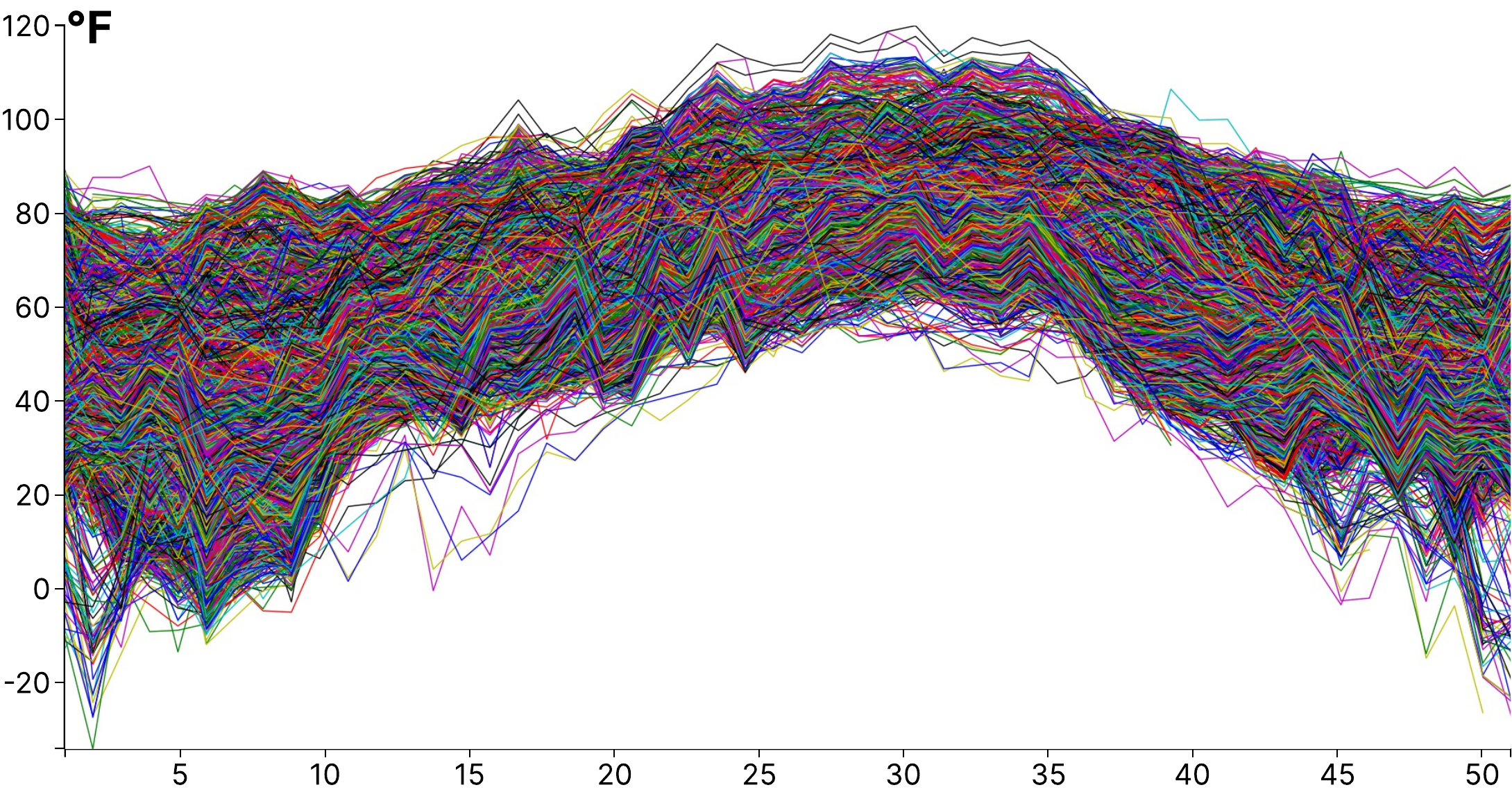}
         \caption{Line-based visualization}
         \label{fig:example-3-raw}
     \end{subfigure}
     \hfill
     \begin{subfigure}[b]{0.32\textwidth}
         \centering
         \includegraphics[width=\textwidth]{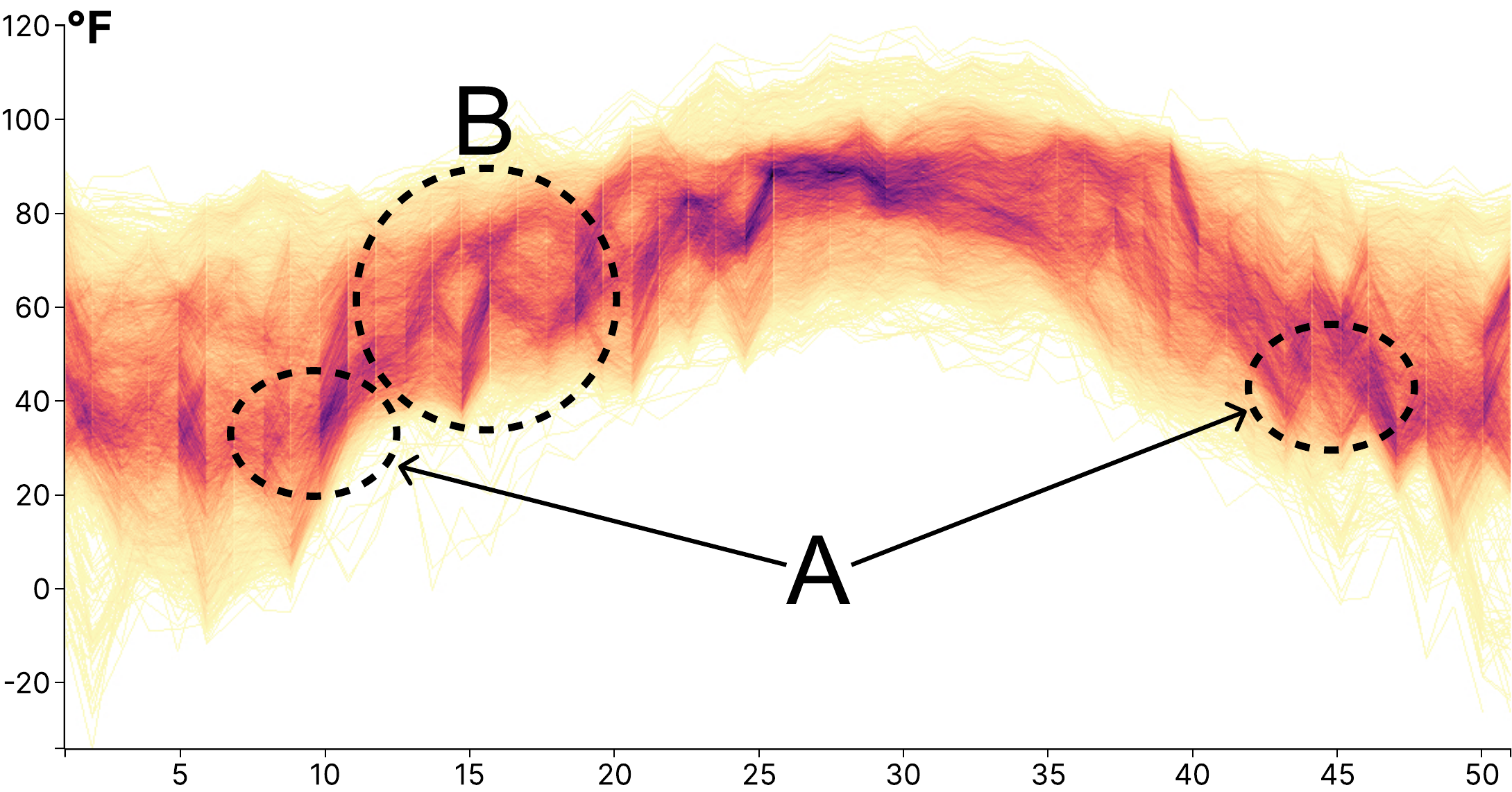}
         \caption{Line-based density plot}
         \label{fig:example-3-density}
     \end{subfigure}
     \hfill
     \begin{subfigure}[b]{0.32\textwidth}
         \centering
         \includegraphics[width=\textwidth]{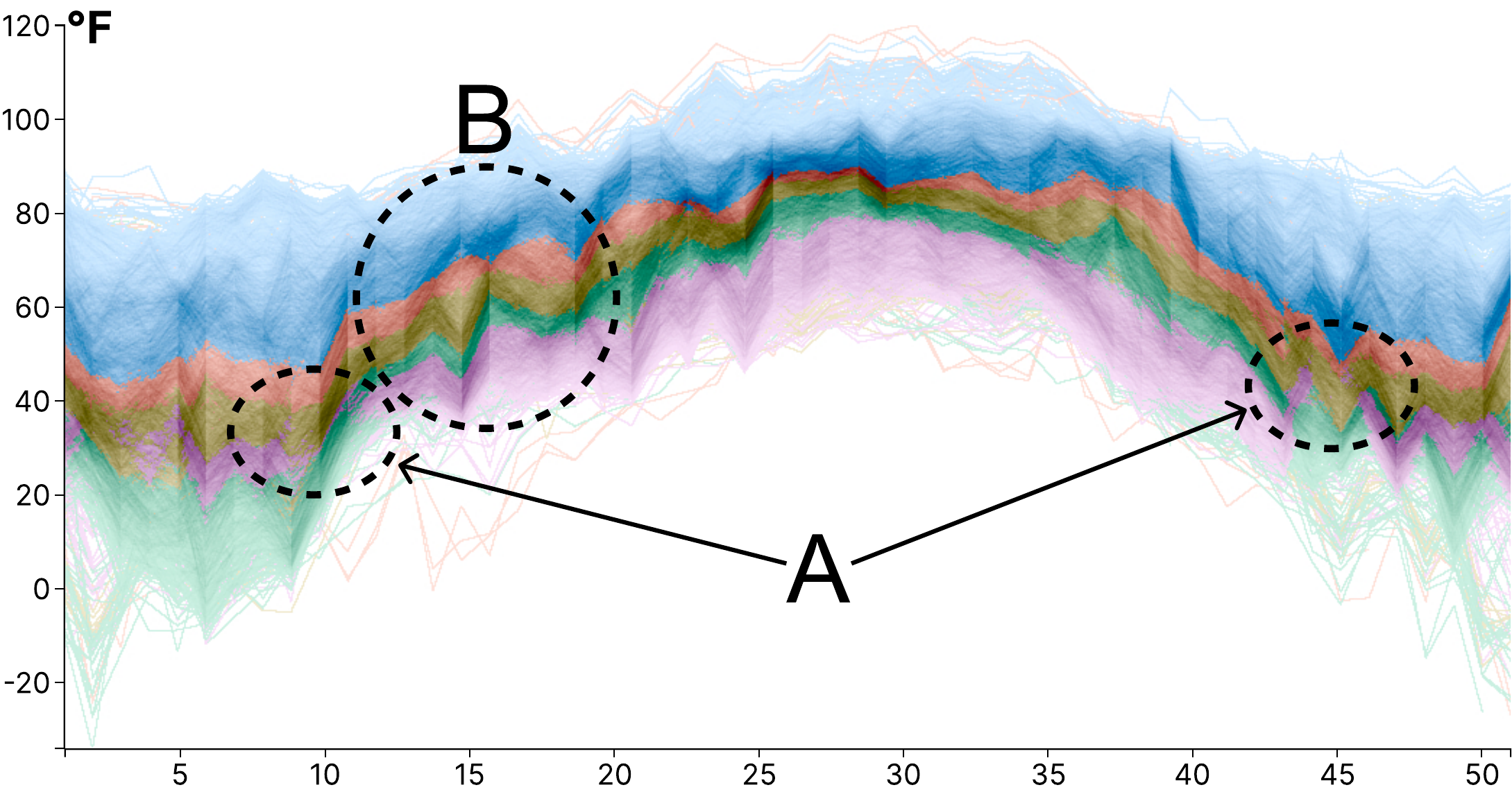}
         \caption{Ours: Colorized density-based plot}
         \label{fig:example-3-colored_marked}
     \end{subfigure}
     \begin{subfigure}[b]{0.32\textwidth}
         \centering
         \includegraphics[width=\textwidth]{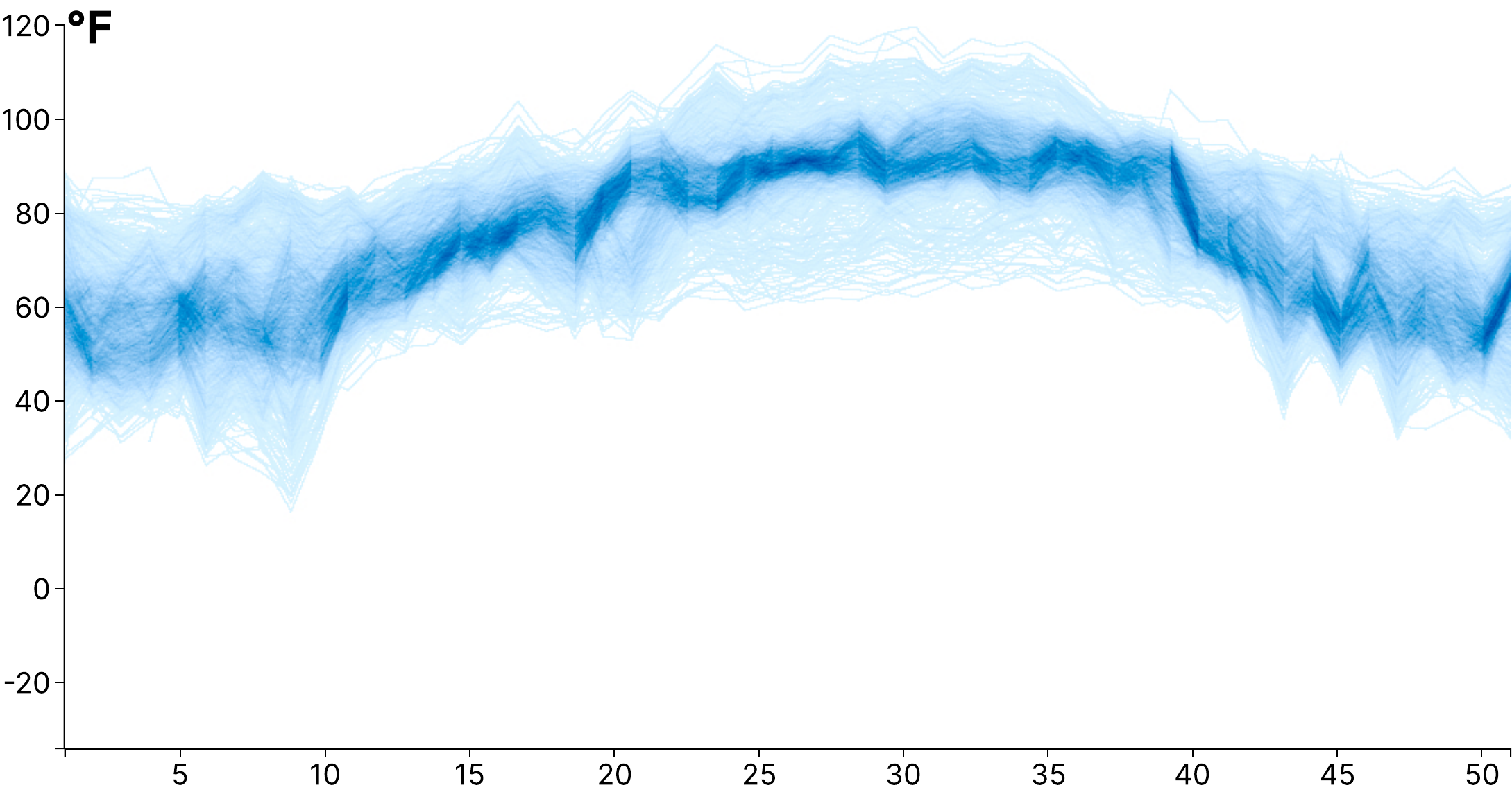}
         \caption{Selected trend with 2180 lines}
         \label{fig:example3_pick1}
     \end{subfigure}
     \hfill
     \begin{subfigure}[b]{0.32\textwidth}
         \centering
         \includegraphics[width=\textwidth]{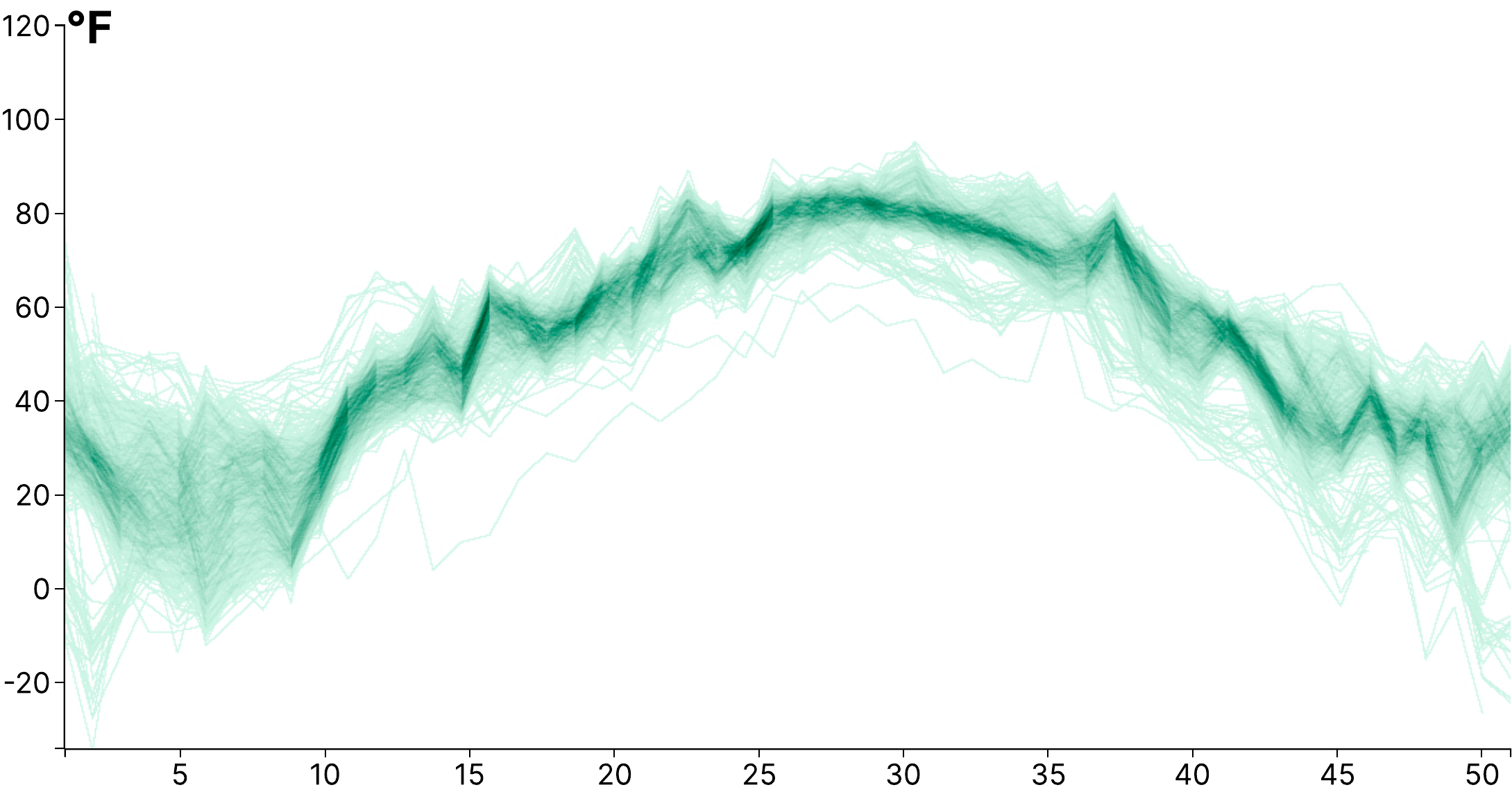}
         \caption{Selected trend with 793 lines}
         \label{fig:example3_pick2}
     \end{subfigure}
     \hfill
     \begin{subfigure}[b]{0.32\textwidth}
         \centering
         \includegraphics[width=\textwidth]{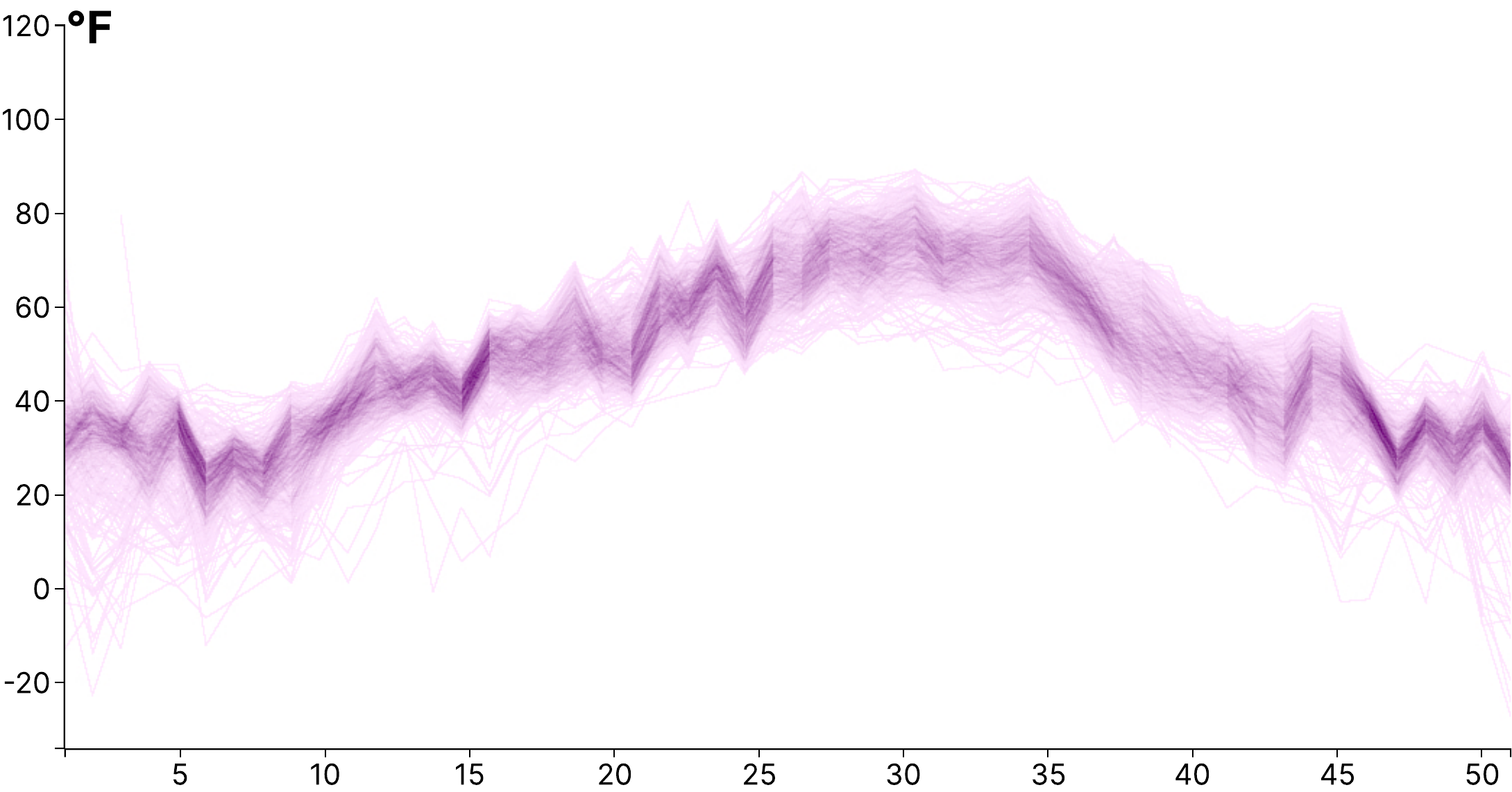}
         \caption{Selected trend with 742 lines}
         \label{fig:example3_pick3}
     \end{subfigure}
    \vspace{-3mm}
    \caption{Time series of temperature values (6187 lines) visualized using (a) line-based visualization, (b) line--based density plot, and (c) our line-based density plot colorization scheme. (d) - (f): Individual trends selected using our interactive tool.}
    \label{fig:usecase_real_world_temerature}
    \vspace{6mm}

    \centering
     \begin{subfigure}[b]{0.32\textwidth}
         \centering
         \includegraphics[width=\textwidth]{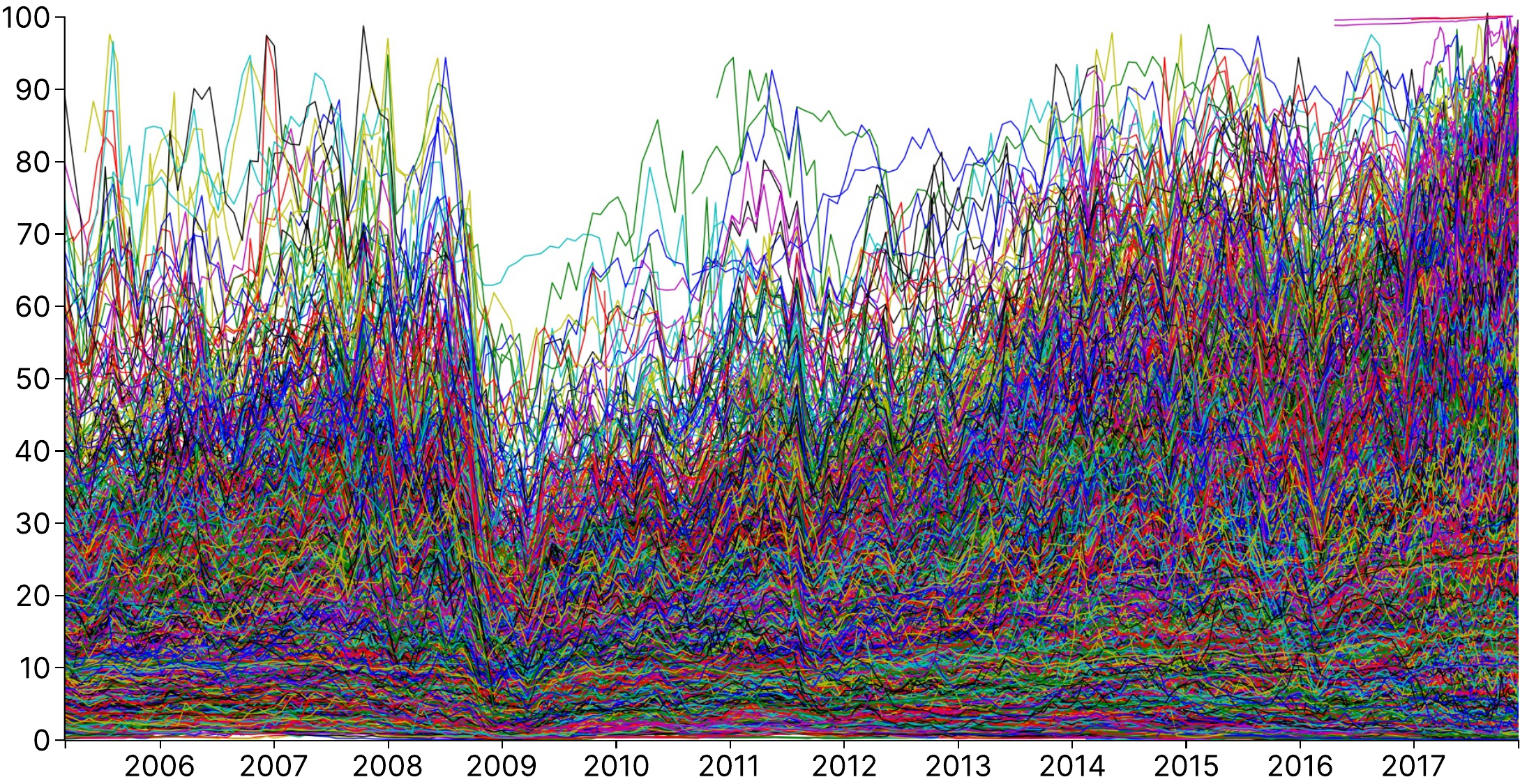}
         \caption{Line-based visualization}
         \label{fig:example-4-raw}
     \end{subfigure}
     \hfill
     \begin{subfigure}[b]{0.32\textwidth}
         \centering
         \includegraphics[width=\textwidth]{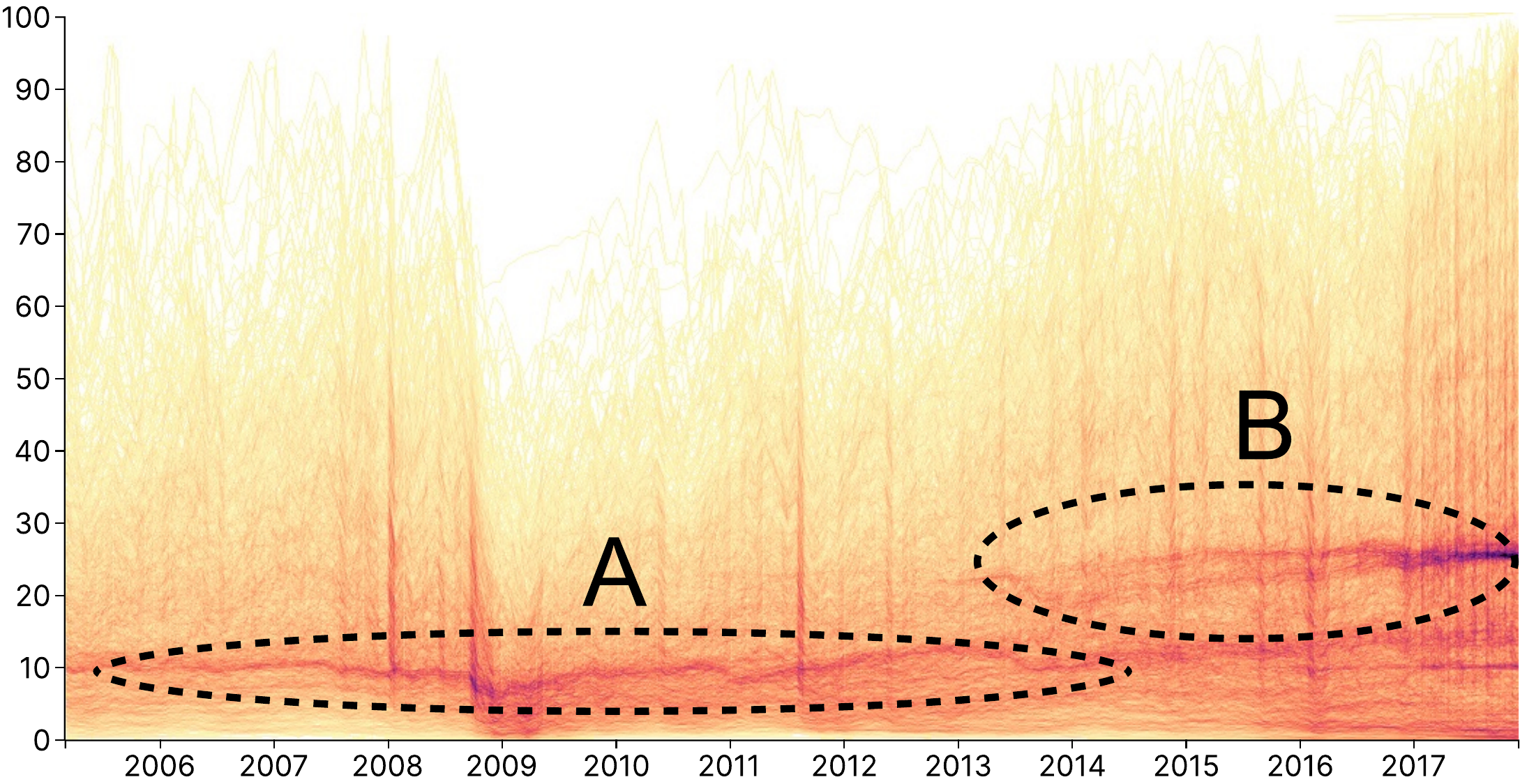}
         \caption{Line-based density plot}
         \label{fig:example-4-density}
     \end{subfigure}
     \hfill
     \begin{subfigure}[b]{0.32\textwidth}
         \centering
         \includegraphics[width=\textwidth]{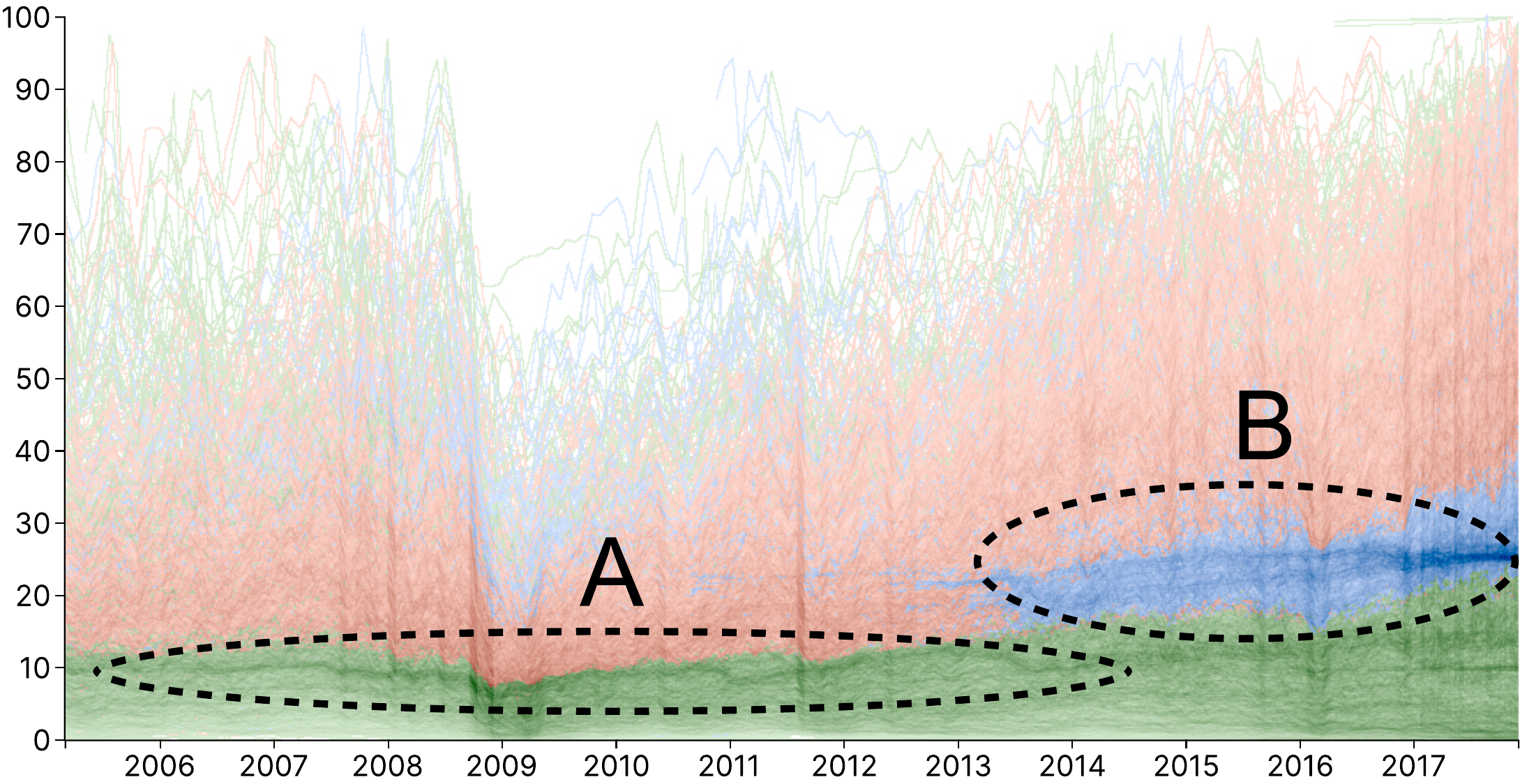}
         \caption{Ours: Colorized line-based density plot}
         \label{fig:example-4-colored_marked}
     \end{subfigure}
     \begin{subfigure}[b]{0.32\textwidth}
         \centering
         \includegraphics[width=\textwidth]{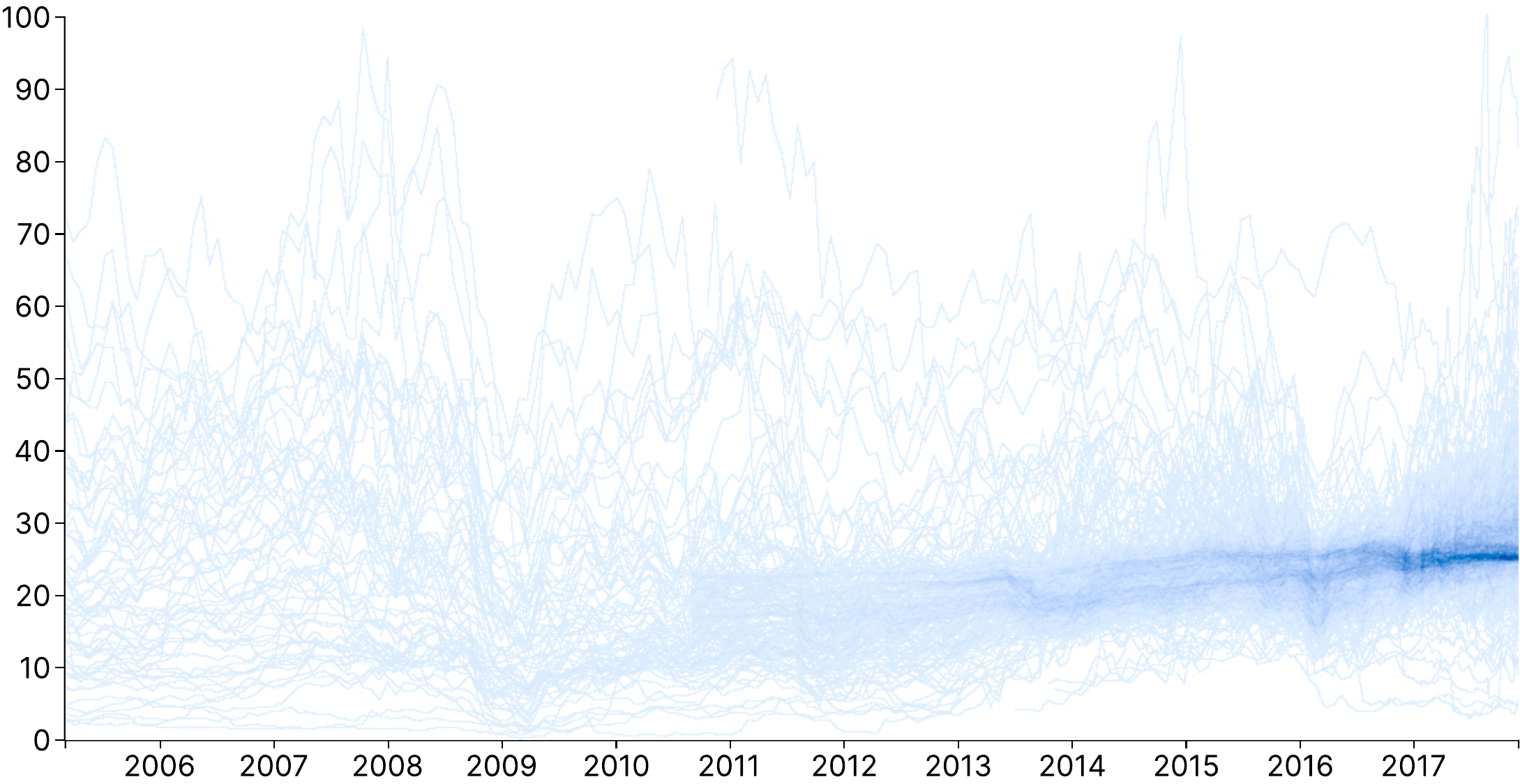}
         \caption{Selected trend with 1034 lines}
         \label{fig:example4_pick1}
     \end{subfigure}
     \hfill
     \begin{subfigure}[b]{0.32\textwidth}
         \centering
         \includegraphics[width=\textwidth]{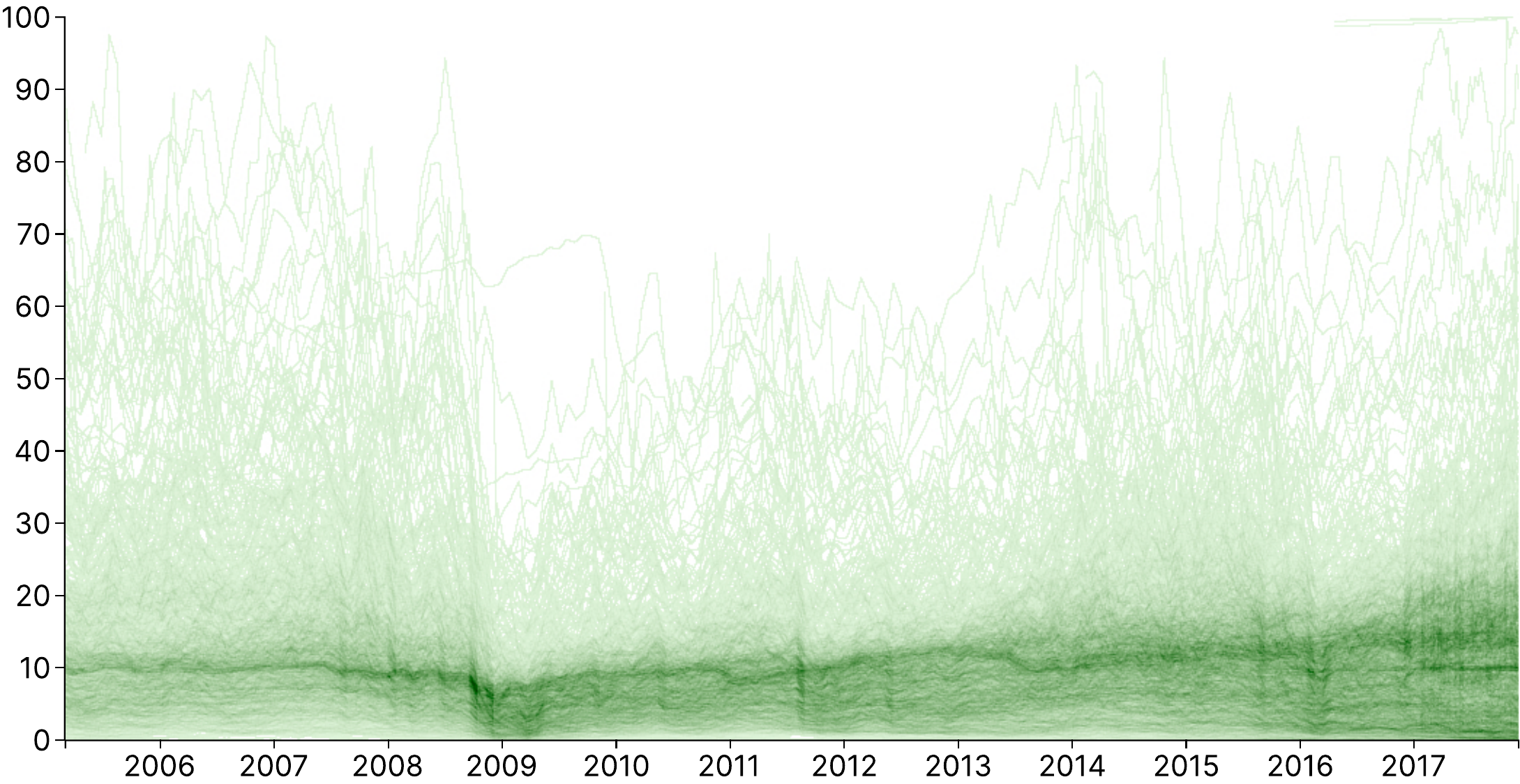}
         \caption{Selected trend with 2368 lines}
         \label{fig:example4_pick2}
     \end{subfigure}
     \hfill
     \begin{subfigure}[b]{0.32\textwidth}
         \centering
         \includegraphics[width=\textwidth]{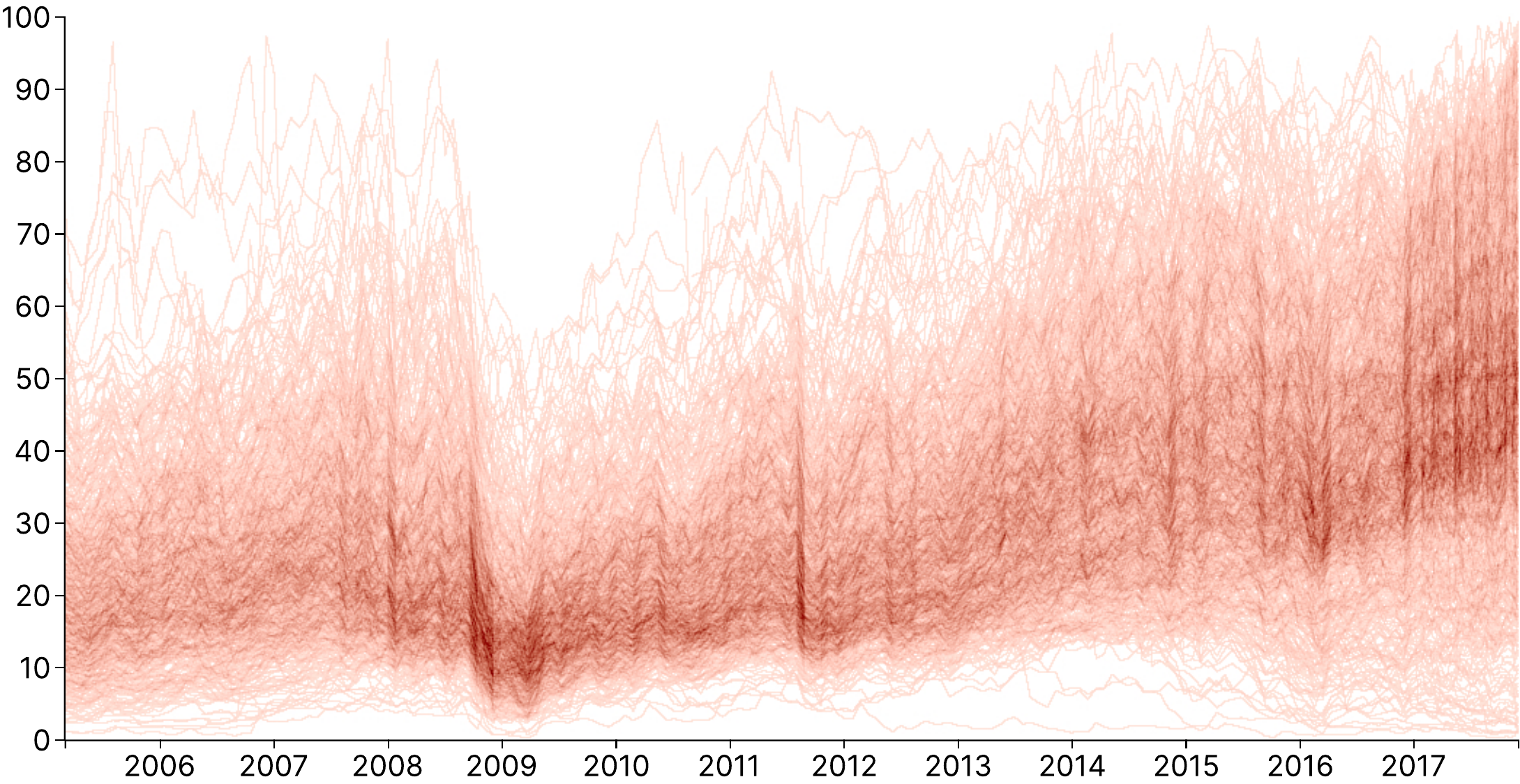}
         \caption{Selected trend with 991 lines}
         \label{fig:example4_pick3}
     \end{subfigure}
    \vspace{-3mm}
    \caption{Daily closing prices of 4393 stock at the New York Stock Exchange from 2005 - 2017 visualized using (a) line-based visualization, (b) line-based density plot, and (c) our line-based density plot colorization scheme. (d) - (f): Individual trends selected using our interactive tool.}

    \label{fig:usecase_real_world_stocks}
    \vspace{6mm}

    \centering
     \hfill
     \begin{subfigure}[b]{0.33\textwidth}
         \centering
         \includegraphics[width=\textwidth]{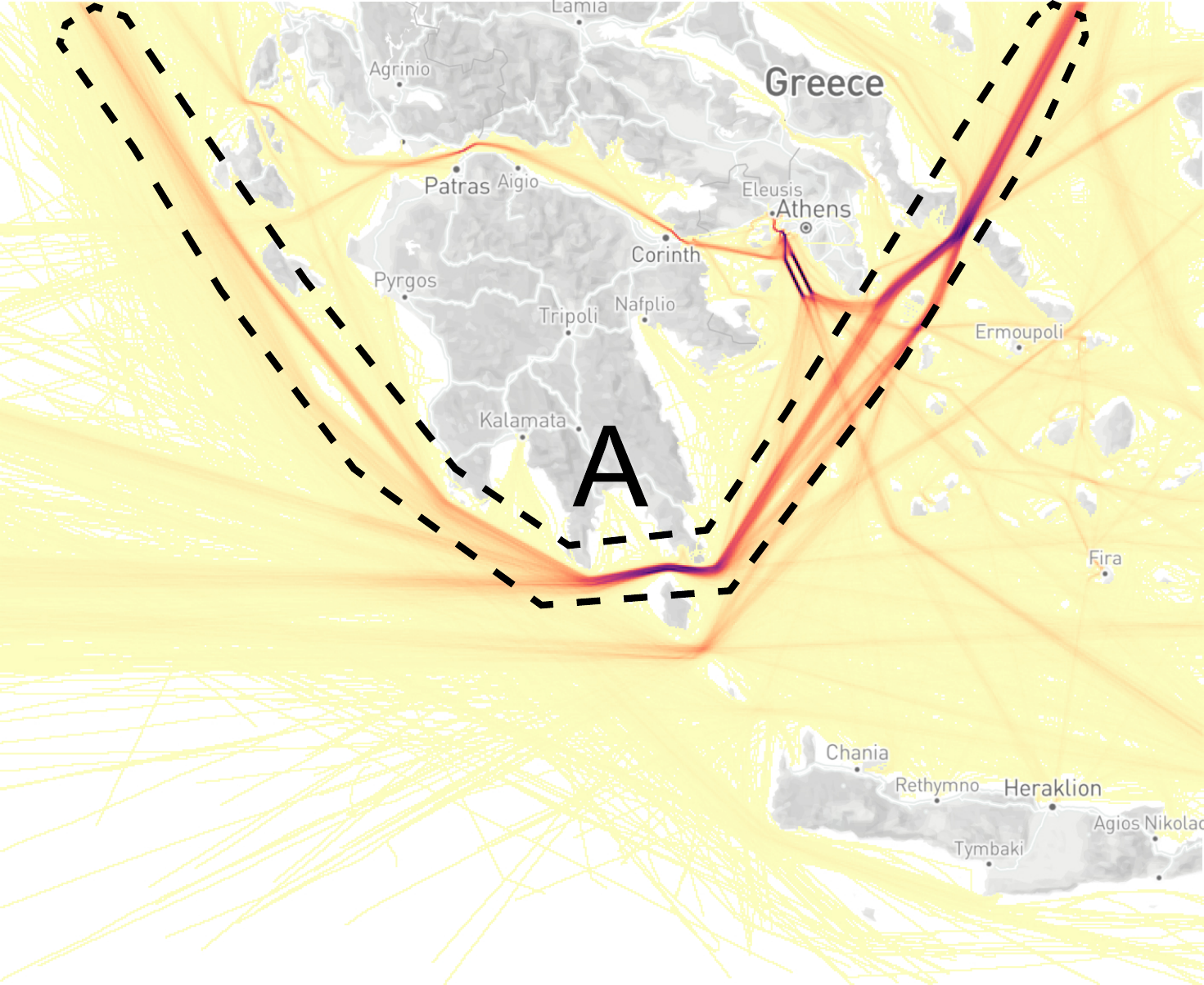}
         \caption{Density-based plot}
         \label{fig:Greece-density}
     \end{subfigure}
     \hfill
     \begin{subfigure}[b]{0.33\textwidth}
         \centering
         \includegraphics[width=\textwidth]{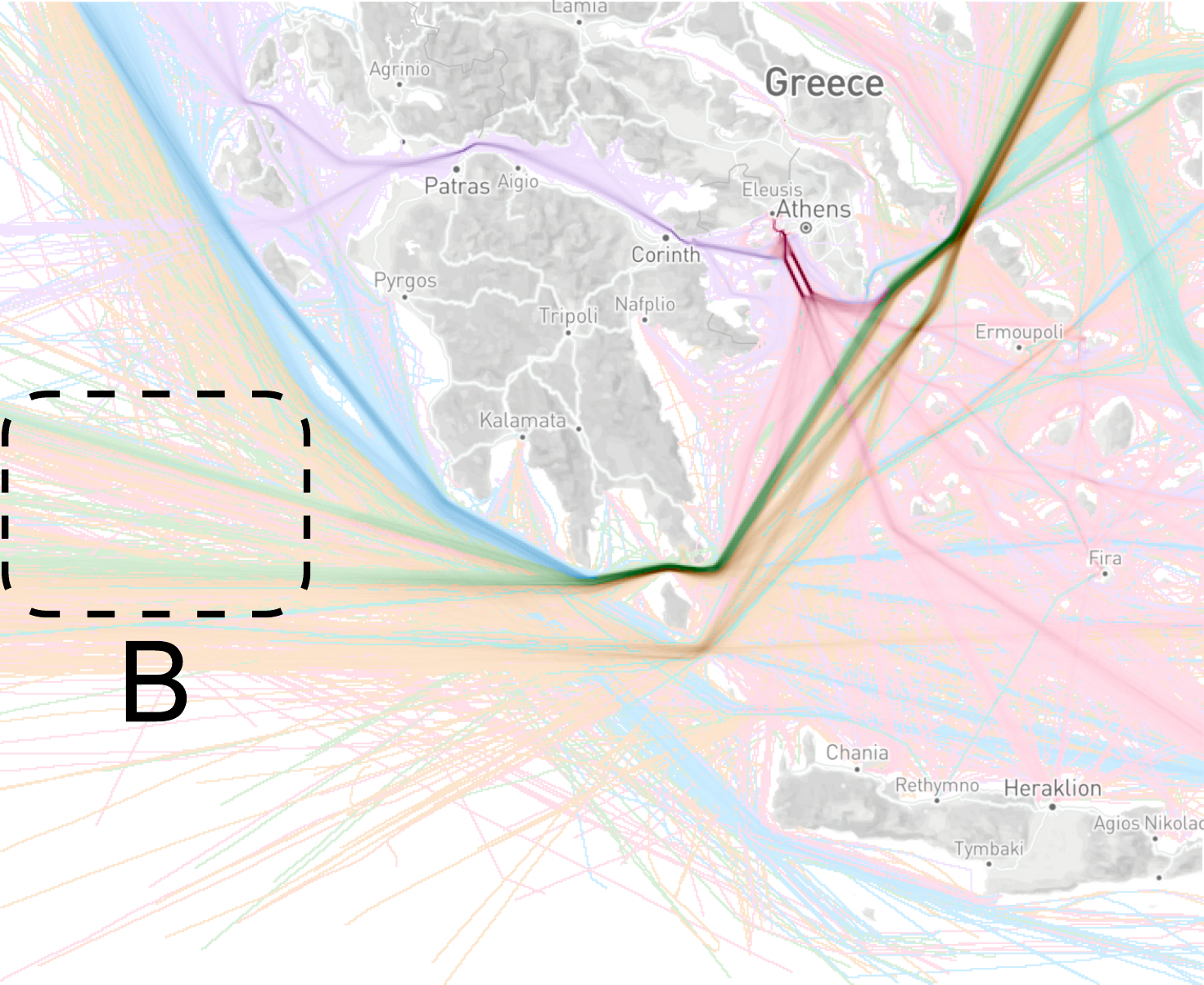}
         \caption{Ours: Colorized density-based plot}
         \label{fig:Greece-colored}
     \end{subfigure}
     \begin{subfigure}[b]{0.33\textwidth}
         \centering
         \includegraphics[width=\textwidth]{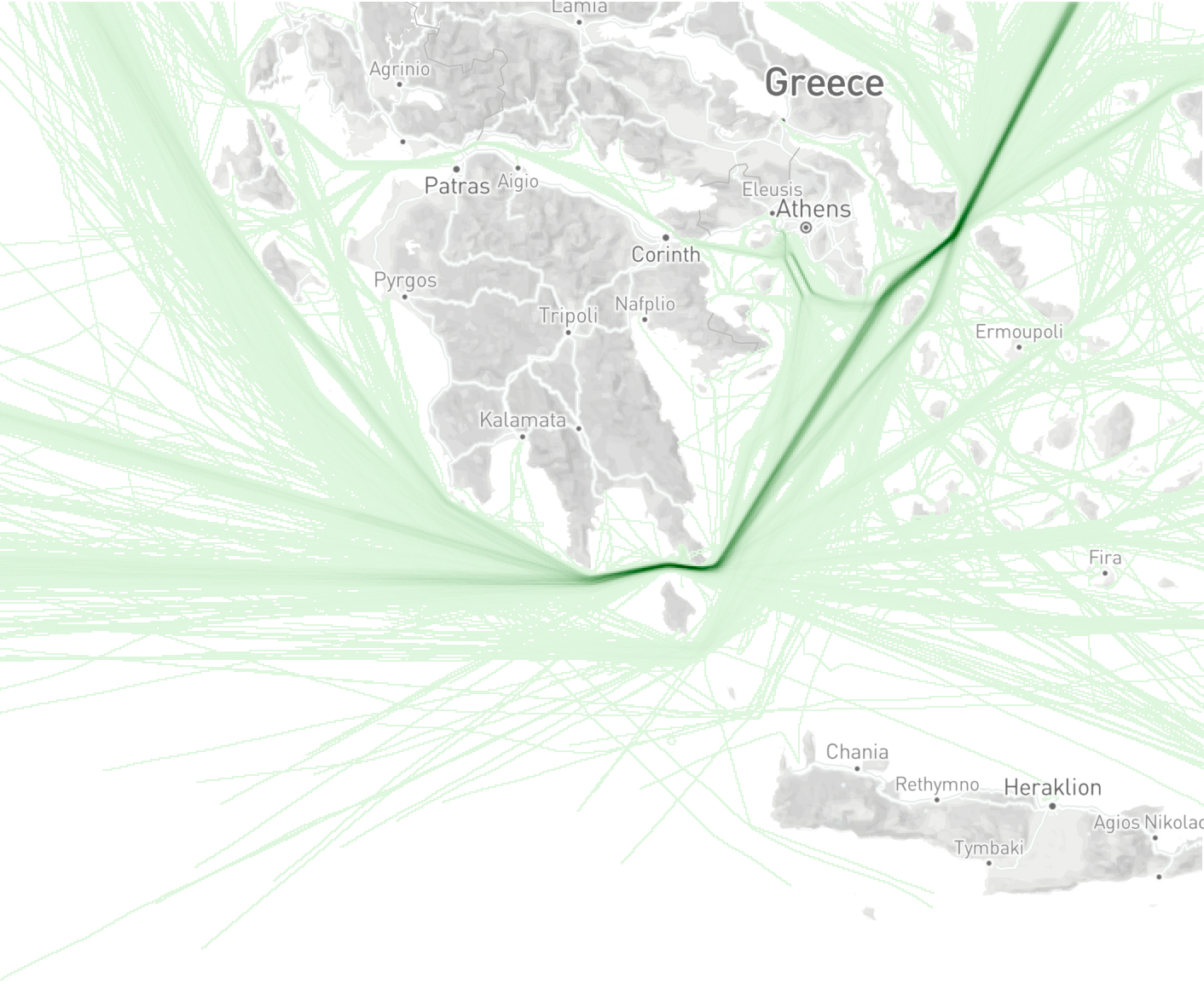}
         \caption{Selected trend with 2604 lines}
         \label{fig:Greece_pick1}
     \end{subfigure}
     \hfill
    \caption{Hellenic Trench AIS data~\cite{frantzis2018hellenic}: (a) density-based visualization of randomly selected 10,000 trajectories with 1,048,575 points. A ``U''-shaped trend seems to dominate the plot; (b) The colored density-based plot instead highlights several high-density trends that intersect; (c) The lines associated with the green cluster show a converging pattern of the cluster after vessels passed the channel (from west to east) between two islands.}
    \label{fig:usecase_Greece}
\end{figure*}

\section{Results}
\label{sec:results}
To demonstrate the effectiveness of our approach and compare it to traditional coloring schemes of density plots. We selected four datasets: two common time-series datasets that have been used in \cite{moritz2018visualizing, zhao2021kd} and two trajectory datasets: Hellenic Trench AIS data~\cite{frantzis2018hellenic} and taxi rides in Beijing~\cite{yuan2010t,yuan2011driving} (featured in the Fig.~\ref{fig:teaser}).

\noindent \textbf{Temperature data.}
This dataset contains 293,175 weekly maximum temperature values of 6187 time series obtained from stations across the United States~\cite{acis}.
The line plot in Fig.~\ref{fig:example-3-raw} is highly convoluted and offers no insight into the underlying trends in the data.
The conventional density plot in Fig.~\ref{fig:example-3-density} visually separates high from low-density regions, but there are ambiguities about how individual trends continue after crossing, e.g., in the region marked ``A'' on the right.
In addition, the parts of the trends circled as ``B'' show zigzag features and align with the ambiguous continuation problem shown in Fig.~\ref{fig:ambiguous_patterns_continuation}.
Our coloring method in Fig.~\ref{fig:example-3-colored_marked} helps to reduce these ambiguities and reveals patterns that are not easily visible otherwise.
In particular, the blue cluster is visually separated from other trends, indicating areas with consistently high temperatures, which makes region ``B'' more clearly stand out. Meanwhile, the purple and green trends intersect at two positions (marked as ``A'' in Fig.~\ref{fig:example-3-colored_marked}), revealing areas that experience sudden temperature changes, making them more anomalous at certain times.
The line counts for these clusters were 2180 (blue, Fig.~\ref{fig:example3_pick1}), 793 (green, Fig.~\ref{fig:example3_pick2}), and 742 (purple, Fig.~\ref{fig:example3_pick1}).
This demonstrates that there are indeed many lines that follow the trend of these clusters. The lines in the blue cluster remain at higher temperatures, while lines of the green and purple clusters cross at certain positions, supporting our previous conclusions drawn from the colored line-based density plot.

\noindent \textbf{Stock market data.}
In the upper row of Fig.~\ref{fig:usecase_real_world_stocks}, we show  552,559 daily closing prices of 4393 stocks (with prices between 0 and 100; we removed stocks with only one close price) at the New York Stock Exchange~\cite{nyse} from 2005 to 2017.
Although the raw line-based visualization in Fig.~\ref{fig:example-4-raw} is cluttered, it is still possible to observe that there are many lines at the bottom of the graph. The pattern (highlighted as ``A'') at the bottom of the density plot in Fig.~\ref{fig:example-4-density} aligns with this observation.
However, there is another high-density pattern, which is shorter and concentrated towards the end of the graph (highlighted as ``B''), that only appears in the density plot.
These two disconnected high-density patterns show the issue of disconnected clusters we presented in Fig.~\ref{fig:ambiguous_patterns_disconneted} in a real-world dataset.
\revised{Our colorization in Fig.~\ref{fig:example-4-colored_marked} illustrates that the clusters marked ``A'' and ``B'' are actually disconnected.} \revised{The cluster ``B''} does not start from the beginning of the time period but instead starts from in between. This is because many stocks were newly listed after 2010 and did not have corresponding records before that time (see Fig.~\ref{fig:example4_pick1}).
After matching lines to the three clusters in the colored line-based density plot, the lines of the individual clusters are shown in the lower row of Fig.~\ref{fig:usecase_real_world_stocks}.
The green cluster represents the main trends and contains to more than half of the total lines (Fig.~\ref{fig:example4_pick2}). It represents a large number of stocks fluctuating at lower prices.
The red cluster contains fewer lines, but the prices of those stocks are relatively high, even though they are more scattered (Fig.~\ref{fig:example4_pick3}).

\noindent \textbf{Ship trajectories.}
The Hellenic Trench AIS dataset~\cite{frantzis2018hellenic}, containing more than 170,000 vessel trajectories, was originally collected to analyze the impact of ship routes on the survival of Mediterranean sperm whales~\cite{frantzis2019shipping}. At first, we cleaned the dataset by removing trajectories crossing land, likely introduced by GPS recording errors. Due to the memory limitation of the browser, we randomly selected 10,000 trajectories with 1,048,575 points to generate the density plot and our colorized version.
The density plot in Fig.~\ref{fig:Greece-density} highlights several intersecting high-density trends.
Visually, a dominant ``U''-shaped trend \revised{(marked as ``A'' in Fig.~\ref{fig:Greece-density})} begins west of the Peloponnese and continues eastwards of Athens.
Similar to Fig.~\ref{fig:ambiguous_patterns_illusionary}, based on the density map alone, it is unclear if the trend is illusory or not. After coloring the density plot by our method (Fig.~\ref{fig:Greece-colored}), the western and eastern parts of the ``U''-shaped trend are colored blue and green. Thus they are likely two separate patterns. Through our line assignment method, 2604 lines were assigned to the green cluster, as shown in Fig.~\ref{fig:Greece_pick1}.
The lines of the green cluster are almost on the same route after entering the channel but beforehand come from various directions.
Although there are some trajectories that match the visual ``U''-shaped trend, such lines are not as dense as indicated in the original density map. Our method succeeds in finding lines that match other clusters and is able to decompose them into several clusters of line bundles.
Another point worth noting is that the pixels in the lower left corner of Fig.~\ref{fig:Greece-colored} \revised{(marked as ``B'')} look like the lines are headed in one direction, but they have been grouped into different clusters. This reflects a limitation of our approach: it is not able to, e.g., measure parallelism or other geometric features of lines.

\begin{figure}[h]
    \centering
   \begin{subfigure}[b]{0.45\linewidth}
         \centering
         \includegraphics[width=\linewidth]{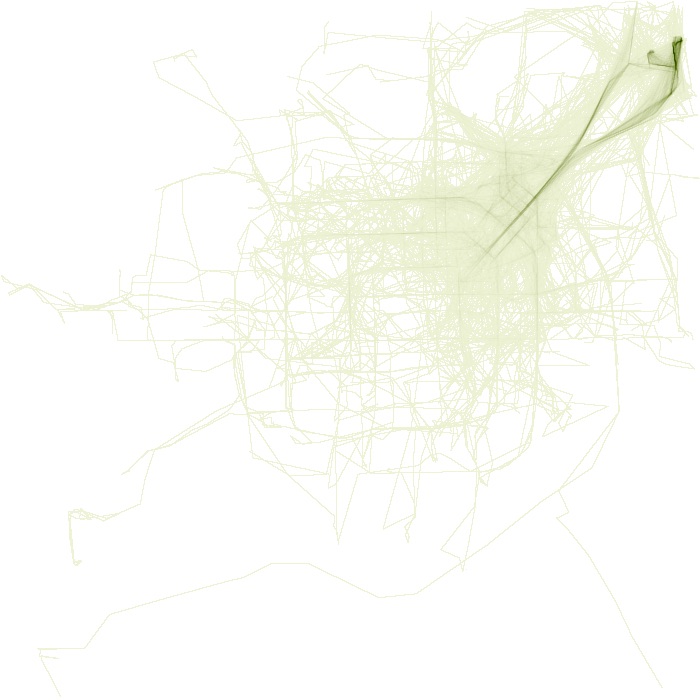}
         \caption{Airport taxi trajectory cluster (1032 lines)}
         \label{fig:teaser_pick1}
     \end{subfigure}
     \hfill
        \begin{subfigure}[b]{0.45\linewidth}
         \centering
         \includegraphics[width=\linewidth]{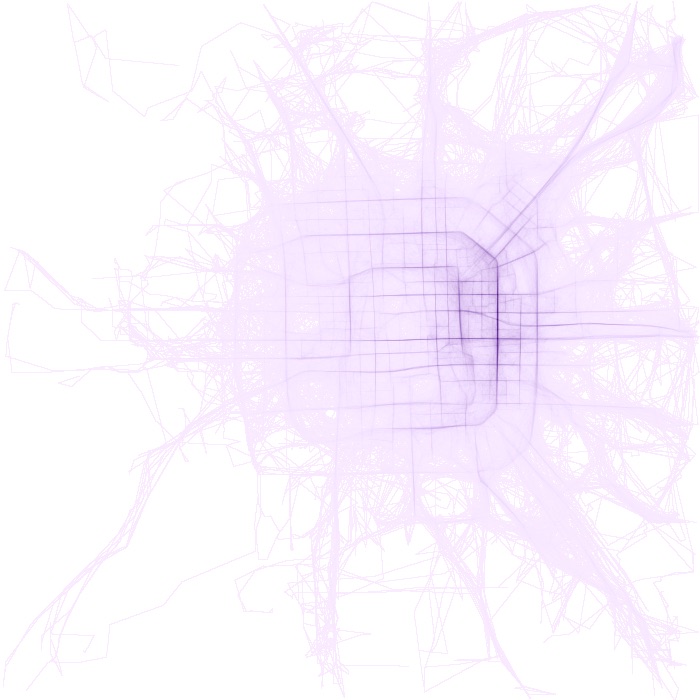}
         \caption{East urban taxi trajectory cluster (1973 lines)}
         \label{fig:teaser_pick2}
     \end{subfigure}
    \vspace{-3mm}
    \caption{Filtered lines from two clusters of Fig.~\ref{fig:taxi_colored}.}
    \vspace{-4mm}
    \label{fig:teaser_picks}
\end{figure}

\noindent \textbf{Taxi trajectories.}
The Beijing taxi trajectory dataset~\cite{yuan2010t, yuan2011driving} contains GPS trajectories of 10,357 taxis on February 2-8, 2008. As this dataset is very large, loading it in our tool exceeded the browser's memory limitations. Thus, we sampled 6502 trajectories of 500 taxis with 734,967 individual time points to generate Fig.~\ref{fig:teaser}.
The line plot in Fig.~\ref{fig:taxi_raw} shows severe overplotting, hiding most of the dataset's structure.
However, although the line-based density plot in Fig.~\ref{fig:taxi_density} reveals the road network of Beijing, it fails to display frequent taxi routes, e.g., it is ambiguous if the taxis circle around the city center or if they drive outwards or inwards from there.
Our method allows to identify different clusters. The lines associated with the green and purple clusters in Fig.~\ref{fig:taxi_colored} are shown in Fig.~\ref{fig:teaser_picks}.
While the green cluster represents taxis driving along the airport highway, where many taxis exit at the toll booths to reach different destinations, the purple cluster represents taxis traveling within the eastern part of Beijing.
This can be explained by the presence of a large residential area east of Beijing.
Similar patterns can be observed for the other clusters. Taxis are unlikely to circle the city center but are likely to drive in one part of the city.

\begin{figure}[tb]
    \centering
   \begin{subfigure}[b]{0.47\linewidth}
         \centering
         \includegraphics[width=\linewidth]{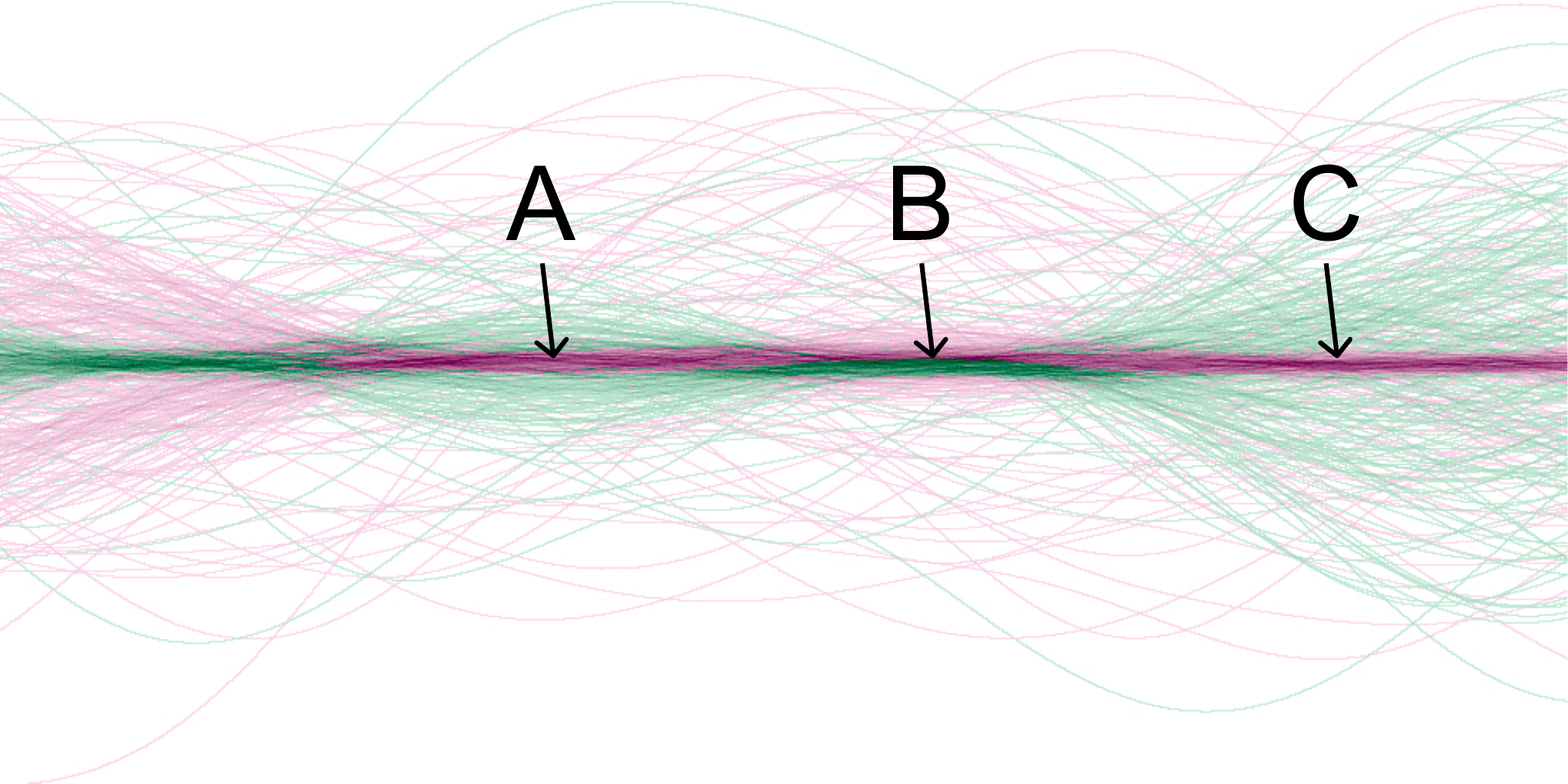}
         \caption{Colored line-based density plot: illusory pattern}
         \label{fig:illusionary_comparison_colored}
     \end{subfigure}
     \hfill
        \begin{subfigure}[b]{0.47\linewidth}
         \centering
         \includegraphics[width=\linewidth]{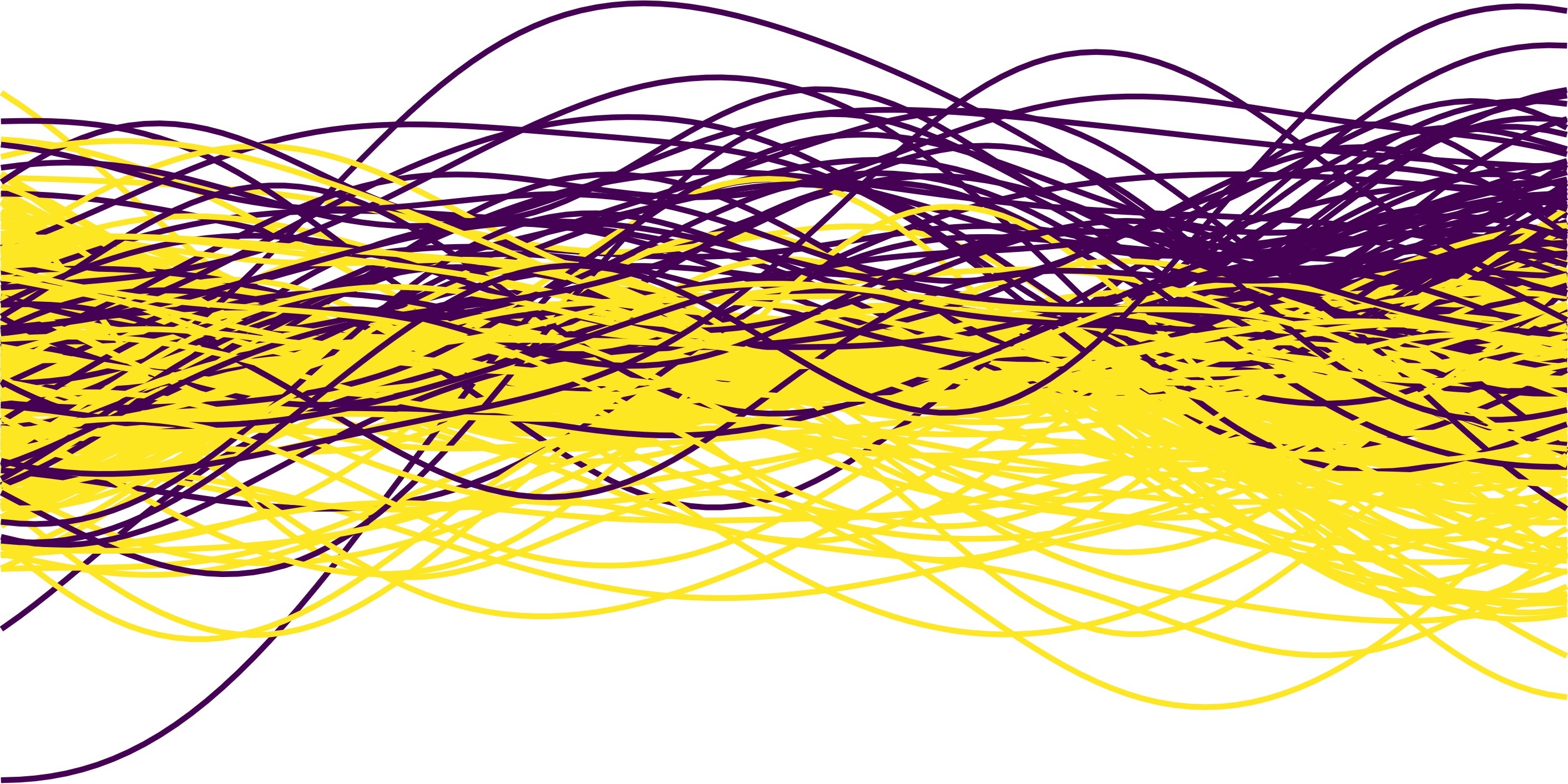}
         \caption{Colored lines: illusory pattern}
         \label{fig:illusionary_comparison_line_colored}
     \end{subfigure}
       \begin{subfigure}[b]{0.47\linewidth}
         \centering
         \includegraphics[width=\linewidth]{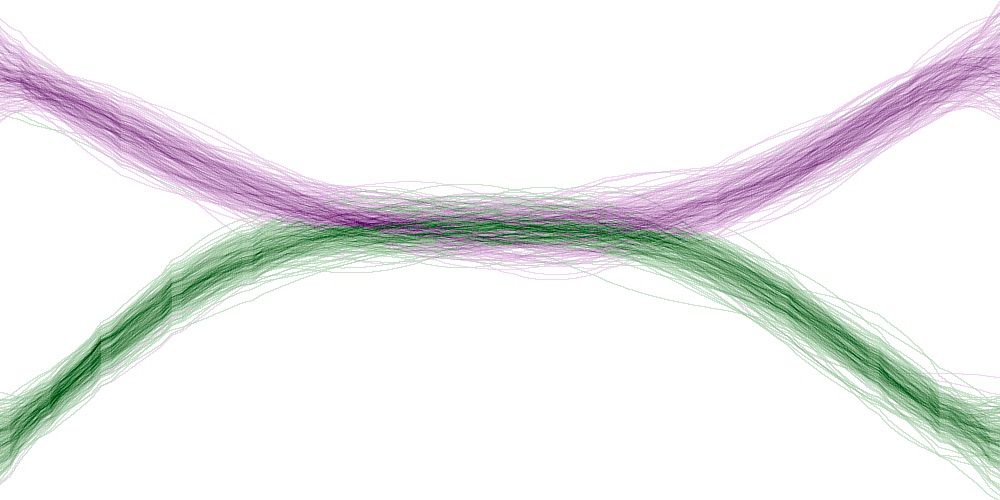}
         \caption{Colored line-based density plot: ambiguous continuation}
         \label{fig:ambiguous_trend_comparison_colored}
     \end{subfigure}
     \hfill
        \begin{subfigure}[b]{0.47\linewidth}
         \centering
         \includegraphics[width=\linewidth]{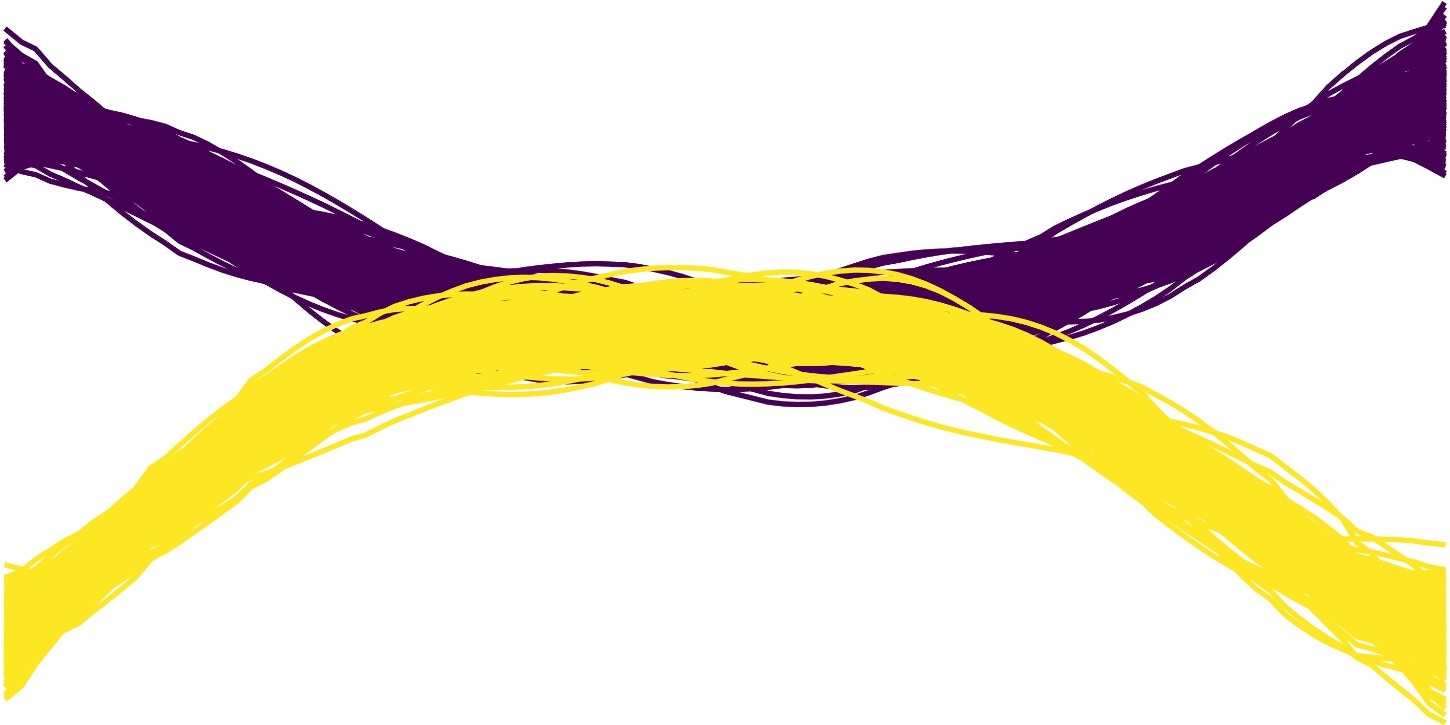}
         \caption{Colored lines: ambiguous continuation \\\hspace{\textwidth}}
         \label{fig:ambiguous_trend_comparison_line_colored}
     \end{subfigure}
     \begin{subfigure}[b]{0.47\linewidth}
         \centering
         \includegraphics[width=\linewidth]{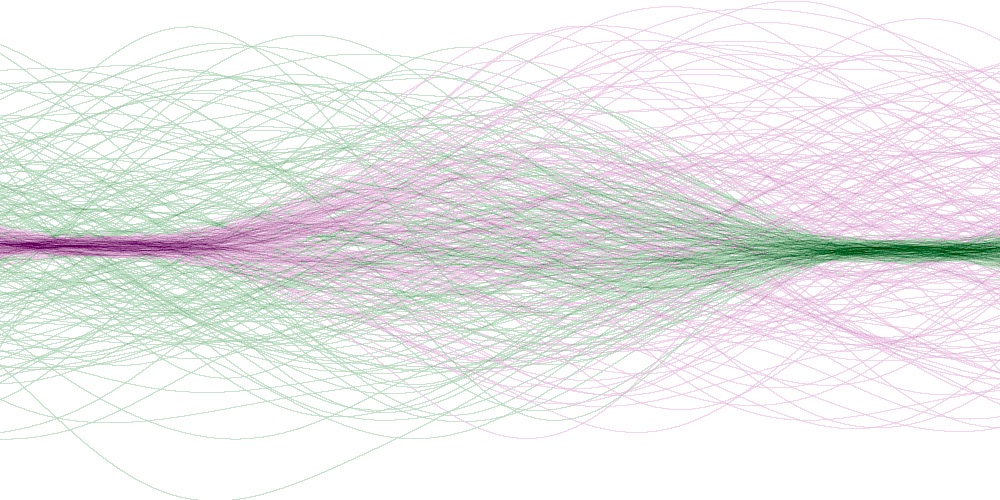}
         \caption{Colored line-based density plot: disconnected clusters}
         \label{fig:disconnected_comparison_colored}
     \end{subfigure}
     \hfill
        \begin{subfigure}[b]{0.47\linewidth}
         \centering
         \includegraphics[width=\linewidth]{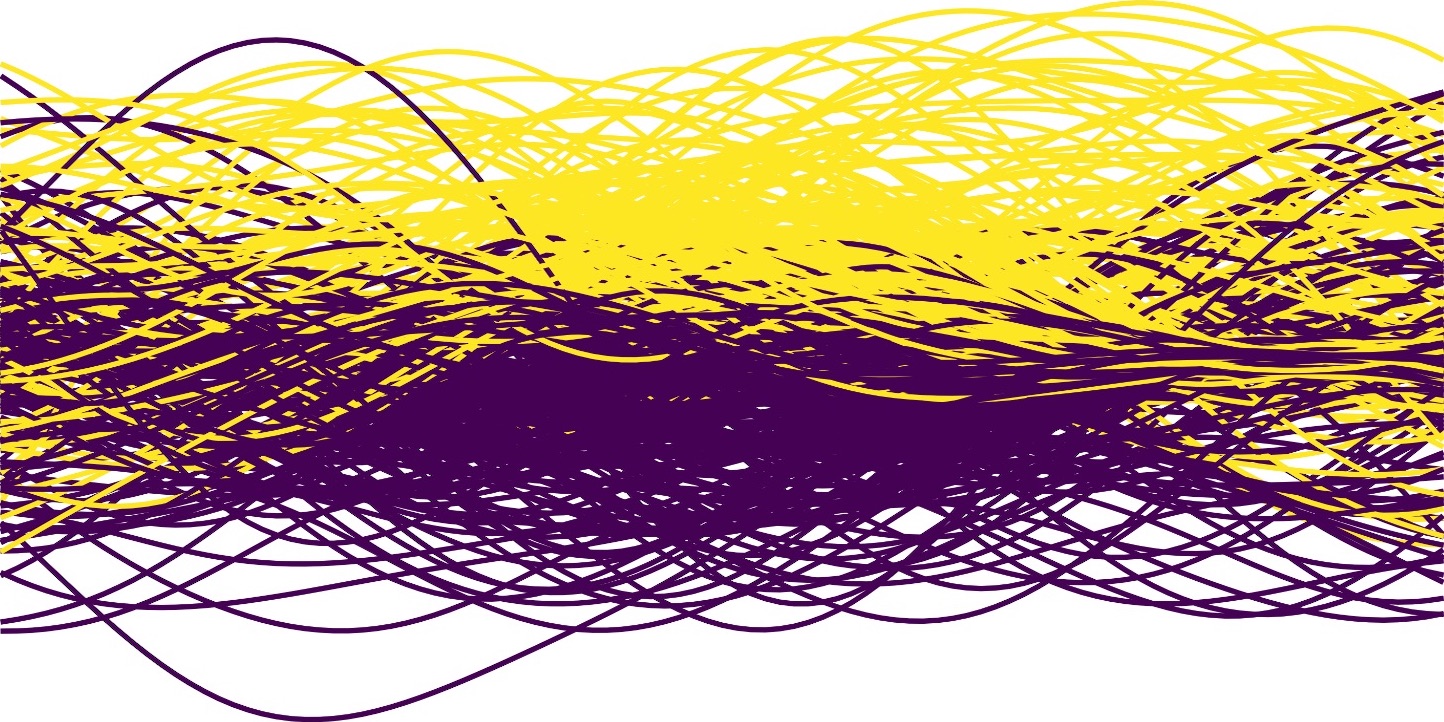}
         \caption{Colored lines: disconnected clusters \\\hspace{\textwidth}}
         \label{fig:disconnected_comparison_line_colored}
     \end{subfigure}
    \vspace{-3mm}
    \caption{Comparison between colorized line-based density plots and line-based clustering for the scenarios of Fig.~\ref{fig:ambiguous_patterns}.}
    \vspace{-6mm}
    \label{fig:comparison_density_line}
    
\end{figure}

\subsection{Comparison to Line Clustering Approaches}
\label{subsec:CompLine}

We use the three synthetic datasets shown in Fig.~\ref{fig:ambiguous_patterns} to illustrate the results of coloring lines by clusters.
The dataset used for Fig.~\ref{fig:ambiguous_patterns_illusionary} contains 400 lines representing the actual pattern and, additionally, 100 lines of background noise. Figs.~\ref{fig:ambiguous_patterns_continuation} and  \ref{fig:ambiguous_patterns_disconneted} contain 200 lines per pattern.

Existing methods treat lines as basic data units and focus on line-based analysis to find trends in the data~\cite{ferstl2015streamline} using clustering. To assign colors, the lines are clustered using a similarity measure tailored to line data.  While these approaches work well for lines with distinct line bundles, they do not work well for noisy data. To highlight the limitations of such methods, we use the state-of-the-art line-based method proposed by Ferstl~{\textit{et al.}}~\cite{ferstl2015streamline} and compare their results to ours.
Their approach involves embedding lines into a higher-dimensional vector space, followed by clustering in a lower-dimensional Euclidean space after using PCA.

To compare our approach to the line-based method, we apply both methods to the examples shown in  Fig.~\ref{fig:ambiguous_patterns}.
While the line-based method fails to distinguish the line bundles in two of the three cases due to a less separable data distribution in low-dimensional space, our approach effectively reveals the underlying structure. 
For instance, in Fig.~\ref{fig:illusionary_comparison_colored}, our method colors the pixels with two colors, showing that the pattern is not a continuous trend (marked as ``A'', ``B'', and ``C'') but is composed of different individual patterns. 
In contrast, the result of the line-based approach in Fig.~\ref{fig:illusionary_comparison_line_colored} cannot reveal the actual components of the trend.
Similarly, in Fig.~\ref{fig:ambiguous_trend_comparison_colored}, our method provides a clear view of the trends of the two line bundles, achieving results that are similar to the line-based method (Fig.~\ref{fig:ambiguous_trend_comparison_line_colored}).
Finally, our approach successfully divides the pixels of the disconnected clusters (Fig.~\ref{fig:disconnected_comparison_colored}), whereas the line-based method (Fig.~\ref{fig:disconnected_comparison_line_colored}) fails to provide valuable information in this case. \revised{Section 5 of the supplementary material, contains further comparative results based on real-world datasets.}

\subsection{Implementation and Interaction}

\noindent \textbf{Implementation.}
Results were obtained using an Apple Silicon M1 processor with 16GB RAM. Our interactive system (\url{https://color-line-density-plot.github.io/}) is written in Javascript and does not utilize the GPU. 
We used Firefox for the runtime tests. The time necessary to calculate the overlap coefficients is dependent on the size of the dataset, resulting in longer computation times for larger sets. For 10,000 lines, our hierarchical tree construction needs about 10s. 
Classifying all bins per division of clusters took less than 4s. The processing speed improves significantly with decreasing cluster sizes. In summary, our method can complete pre-processing (including threshold filtering, sampling, and hierarchical tree construction) for all datasets presented in this paper within 20s. Interactions to separate clusters took less than 4s, mostly between 1s and 2s.

\noindent \textbf{Density threshold and sampling rate.}
Section \ref{sec:sampling} outlines the sampling strategy that requires setting density threshold and sampling rate. 
To enhance the intuitiveness of the minimum threshold, bins below the threshold are hidden using a corresponding slider. 
The user is informed when the selected bins exceed the processing capacity.

\noindent \textbf{Cluster operations.}
Users can set the number of clusters as well as interactively separate clusters. After setting a number, clusters will be divided accordingly, based on the distance between the nodes of the hierarchical tree. Clicking on a cluster allows the user to split it into two sub-clusters. The feature vectors of these sub-clusters will then be mapped using circular MDS to find appropriate hue values. Users can select cluster colors that are fixed in the hue circle.
Then, during gradient descent, only unfixed points are moved.

\noindent \textbf{Hue adjusting and harmonization.}
In Section~\ref{subsec:coloring}, we describe our use of multidimensional scaling (MDS) to ensure that the distance relationships between hues within each cluster correspond to their distance relationship in high-dimensional data space. 
Nonetheless, users can modify the hue by dragging the dot representing the cluster on the hue wheel.
Additionally, we provide users with eight harmonic templates, as defined by Cohen-Or~{\textit{et al.}}~\cite{cohen2006color}.
Once a template is selected, the hues for each cluster will be mapped to the corresponding area, with the movement of the dots being constrained accordingly.
\section{Motivational User Study}
\label{sec:user-study}
Here, we present the findings of a small motivational user study we performed to understand the ability of users to identify and trace trends in line-based density plots.
We explored whether users have issues separating trends of lines.
Additionally, we investigated if users are even aware of potential ambiguities in given plots. 
We did not include a comparison between our proposed colorization method and traditional density plots.
We expected participants to perceive different trends overall, but as long as the \revised{line-based density plot} remains \revised{visually similar, even if it is generated based on different underlying data, their perceptions should remain unchanged.}
We posed the following two hypotheses: \\
\textbf{H1:} There are different understandings when participants perceive trends in ambiguous line-based density plots.\\
\textbf{H2:} The participants are mostly unaware of the ambiguities in line-based density plots.

\subsection{Experimental Design}
We conducted a  browser-based online user study with university students and Ph.D. candidates.
A total of ten line datasets were included, all presented in the form of line-based density plots.
Participants had to identify trends in line-based density plots and complete three tasks for each dataset. \revised{All plots used in the user study and the trends they are composed of are available in Section 3 of the supplementary material.}

\noindent \textbf{Tasks:}
\begin{itemize}
\setlength{\itemsep}{0pt}
\setlength{\parsep}{0pt}
\setlength{\parskip}{0pt}
    \item \textit{Tracing task.}  The participants were asked to trace each \revised{dominant} trend they identified.
    \item \textit{Counting task.} Participants were asked to count the number of trends they perceived for each line-based density plot.
    \item \textit{Certainty task.} We let participants rate how certain they were about their judgments on the number and tracing of trends.
    
\end{itemize}

\noindent \textbf{Ambiguous pattern generation.}
We generated ten line datasets in total based on the three types of ambiguities introduced in Section \ref{sec:illusorypatterns}. 
For each kind of ambiguous pattern, we provided multiple examples with varying degrees of ambiguity or different numbers of trends. 
In the first three datasets, we used illusory patterns, similar to Fig.~\ref{fig:ambiguous_patterns_illusionary}.
The line-based density plot of the first dataset was composed of a horizontal trend superimposed on noise, which is similar to the right trends in the figure.
For the following two datasets, we combined trends like the ones on the left in the top row of the figure. We increased the fan out in the center of each individual trend. Thus the plots were getting progressively less ambiguous, as seen in Fig.~\ref{fig:user_study_number_distribution_illusionary_2}.
As the second kind of ambiguous pattern, we added ambiguous continuations to datasets 4 to 6.
The combined patterns from Fig.~\ref{fig:ambiguous_patterns_continuation}, both the upper left and the upper right composition, are used for datasets 4 and 5. 
In addition, we added another line dataset whose density
plot was composed of three individual trends (dataset 6).
The third kind of pattern showed disconnected clusters.
In the last four datasets, the density plots contained potentially disconnected clusters.
datasets 7 and 8 were composed of distinct patterns.
On one hand, dataset 7 consisted  of a continuous trend with a central fan-out region superimposed on low-density noise, as shown on the right side of the upper row in Fig.~\ref{fig:ambiguous_patterns_disconneted}.
On the other hand, dataset 8 comprised two independent trends, which are illustrated in Fig.~\ref{fig:user_study_number_distribution_disconnected_1}. 
To decrease the ambiguity of the pattern, we increased the horizontal length of the high-density regions of the two independent trends to create datasets 9 and 10. This reduced the size of the central overlapping region as shown in Fig.~\ref{fig:user_study_number_distribution_disconnected_2}.

\noindent \textbf{Experiment organization.}
We displayed the line-based density plots of the datasets colored using a multi-hue color map, one dataset per page, on our survey website. 
All participants had to fulfill the three tasks for each of the ten datasets in the same order.
For the datasets with illusory patterns (datasets 1 to 3), because we were interested in whether participants could tell if multiple trends were combined, we asked them to trace only the main trends.
For the datasets with ambiguous continuation (datasets 4 to 6), we asked participants to trace the individual trends they perceived with different color. 
For the disconnected clusters  (datasets 7 to 10), we asked participants to mark the trends they thought were connected with the same color of strokes, and if they thought they were disconnected, use different colors.

\noindent \textbf{Participants.}
In total, we collected inputs from 25 participants who completed our user study effectively. Three other participants were excluded as they either did not trace any trends or entered numbers of perceived trends that vastly differed from the trends they actually drew. We gathered the traces for all ten tasks drawn by each participant. The participants' age was, on average, 26 years with a standard deviation of 4. The gender distribution among our participants was 56\% male, 36\% female, and 8~\% preferring not to tell their gender. 

\noindent \textbf{Procedure.}
Each participant went through the following procedure: 
(1) completing a short questionnaire with questions about age, gender, and familiarity with line charts and density plots;
(2) viewing an introduction about the tasks based on an already marked density plot; and
(3) performing the tasks for each dataset.

\begin{figure}
    \centering
    \resizebox{\columnwidth}{!}{\import{figs/}{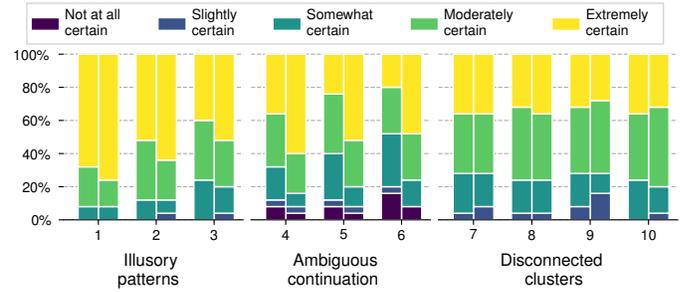}}
    \vspace{-5mm}
    \caption{Responses to the Likert scale of 25 participants for 10 ambiguous line-based density plots. The left bars represent the participants' certainty about the number of trends they identified. The right bars represent the participants' certainty about the trends they traced. }
    \label{fig:user-study}
    \vspace{-6mm}
\end{figure}
\subsection{Experimental Results}
\begin{figure}
\begin{minipage}{0.3\linewidth}
        \begin{subfigure}{\linewidth}
    \includegraphics[width=\linewidth]{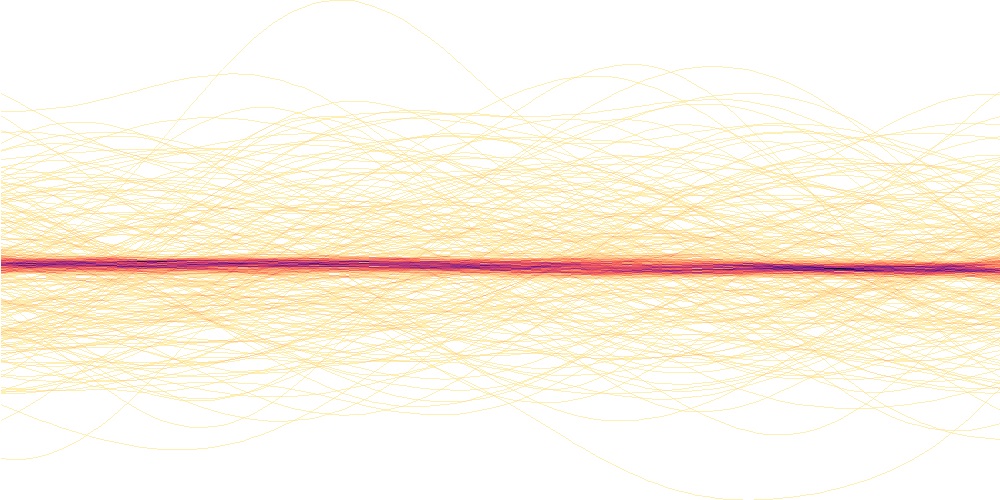}
        \caption{}
        \label{fig:user_study_number_distribution_illusionary_1}

        \end{subfigure}
        \begin{subfigure}{\linewidth}
    \includegraphics[width=\linewidth]{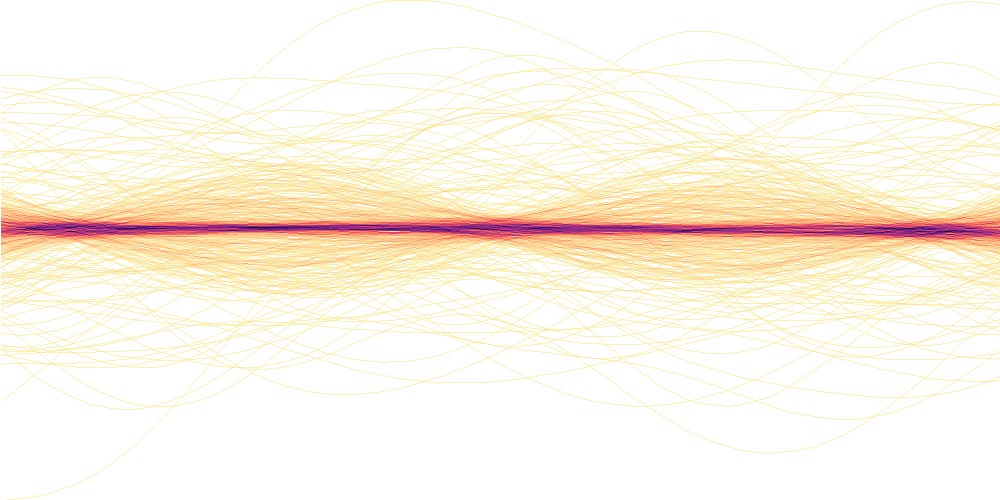}
        \caption{}
        \label{fig:user_study_number_distribution_illusionary_2}

        \end{subfigure}
        \end{minipage}
        \hfil
        \begin{minipage}{0.38\linewidth}
        \begin{subfigure}{\linewidth}
    \resizebox{\linewidth}{!}{\import{figs/}{histogram2_reorder.pgf}}
    \caption{}
    \label{fig:bar_chart}

        \end{subfigure}
         \end{minipage}
         \begin{minipage}{0.3\linewidth}
        \begin{subfigure}{\linewidth}
    \includegraphics[width=\linewidth]{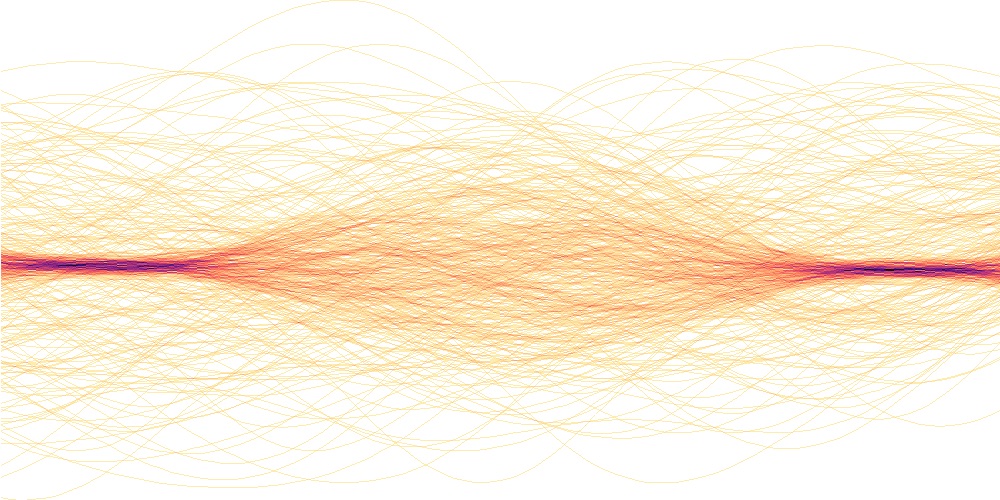}
        \caption{}
        \label{fig:user_study_number_distribution_disconnected_1}
        \end{subfigure}
        \begin{subfigure}{\linewidth}
            \includegraphics[width=\linewidth]{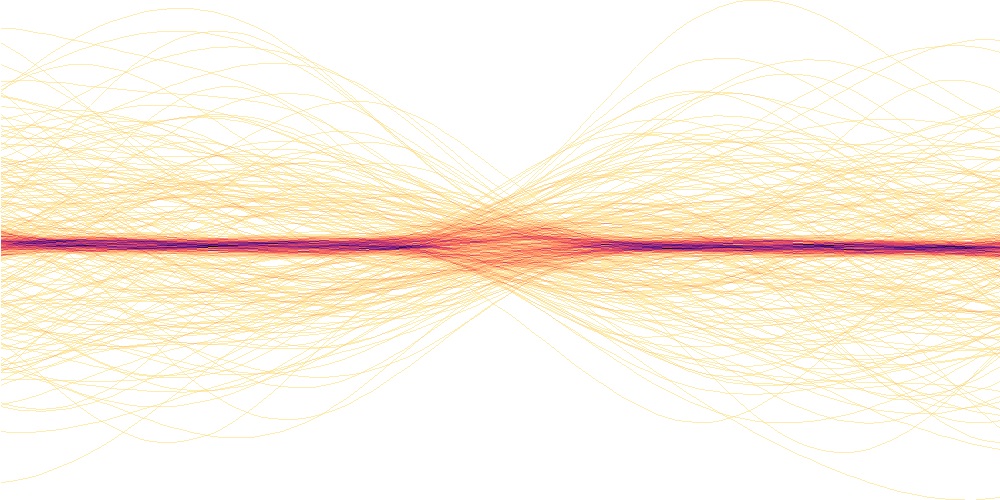}
        \caption{}
        \label{fig:user_study_number_distribution_disconnected_2}
        \end{subfigure}
        \end{minipage}
            \vspace{-3mm}
    \caption{Line-based density plots with illusory trends with varying fan-out of lines from (a) to (b) and disconnected clusters with varying separation from (d) to (e). (c) Distribution of the number of perceived trends in \revised{10} line-based density plots.}
    \label{fig:user_study_number_distribution}
    \vspace{-6mm}
    \end{figure}

Fig.~\ref{fig:user-study} illustrates the participants' responses using a Likert scale to the questions about their certainty regarding the number of trends and the tracings of the trends. 
Fig.~\ref{fig:user_study_number_distribution} shows the distribution of the number of trends the participants perceived for each line-based density plot.

\noindent \textbf{Illusory patterns.}
Most participants perceived a central horizontal trend in the density plots of the illusory patterns shown in Fig.~\ref{fig:user_study_number_distribution_illusionary_1}.
The first dataset, consisting of only a horizontal trend superimposed by noise, was correctly identified by all participants and traced using a horizontal line.
For all datasets with illusory patterns, more than 70\% participants stated that they were ``Moderately certain'' or ``Extremely certain'' about the number of trends and their tracings. However, increasing the fan-out region of the combined trends in the illusory pattern, as shown in Fig.~\ref{fig:user_study_number_distribution_illusionary_2}, decreased the participants' certainty about how many trends they perceived and how they traced them. Fig.~\ref{fig:user-study} shows a decrease in the portion of participants being ``Extremely certain'' about the number of trends they identified and how they traced them from the first dataset to the third. Additionally, the portion of participants perceiving at least two trends increased from 0\% to 16\%, as shown in Fig.~\ref{fig:user_study_number_distribution}. 
This indicates that participants were increasingly unsure about the trends in the density plots.
Still, even though datasets 2 and 3 were created from data containing multiple trends, participants mostly perceived only a single trend for all datasets (92\% and 84\%, respectively), which supports \textbf{H2}.\\
\noindent \textbf{Ambiguous continuation.}
All participants correctly identified, with moderate to high certainty, that the combined line-based density plots of the individual ambiguous pattern shown in Fig.~\ref{fig:ambiguous_patterns_continuation} are composed of two trends.
For dataset 6, consisting of three trends, all participants except for one correctly identified the number of trends.
In the tracing task, 88\% of the participants identified the same trends in both datasets 4 and 5. 
60\% of the participants traced the combination of two ``U''-shaped trends in both plots.
While 28\% of the participants traced crossing trends for both plots in an ``X'' shape.
For datasets 4 and 5, whose density plots are almost identical, we expect participants to perceive them ambiguously. \revised{For dataset 4, 16 participants identified a ``U''-shaped pattern and 9 an ``X''-shaped. For dataset 5, 17 participants traced a ``U'' and 8 an ``X''.}
For dataset 6, which contains a combination of three trends, the participants' understandings are more diverse.
\revised{
We only offered the ability to trace a single trend in the data. Participants were aware that there might be multiple possible trends in the plot but drew only the most dominant one. Therefore, the results do not show whether participants potentially perceived multiple patterns. The distribution of participants among both identified trends indicates that there was no consensus on which trend was dominantly perceived. We conclude that this supports \textbf{H1} with respect to the group of participants, but not for an individual participant.
}\\
\noindent \textbf{Disconnected clusters.}
While participants largely agreed on the number of perceived trends in the previous patterns, they were torn about whether the disconnected clusters in Fig.~\ref{fig:ambiguous_patterns_disconneted} contained one or two trends.
For dataset 7, 56\% and for dataset 10, 64\% of the participants indicated that the trends were independent, like in the left example of Fig.~\ref{fig:ambiguous_patterns_disconneted}.
The remaining portion of participants highlighted just one continuous trend, as shown on the right.
Thus participants did not agree, while over 80\% of them were at least moderately certain about the number of patterns they identified and their tracings.
Fig.~\ref{fig:bar_chart} shows the disagreement of the participants regarding the number of trends for datasets 7-10.
As the density plots of datasets 7 and 8 are almost identical but contain different trends, we expect participants to give similar answers.
For dataset 7, 44\% of the participants answered one trend, and 56\% of the participants answered two.
While for dataset 8, it is 40\% and 60\%, respectively.
For datasets 9 and 10, the participation response percentages are still similar, although they are less ambiguous.
This further supports \textbf{H1}.

\subsection{Experimental Findings}
While our user study was meant to exemplify the ambiguity in the user's understanding of line-based density plots, it is not a large-scale study with extensive possibilities for statistical evaluations.
Nevertheless, it supports our hypothesis \textbf{H1} and \textbf{H2} and shows trends in line-based density plots are perceived ambiguously. 
While participants can correctly identify the number of trends for the patterns that showed ambiguous continuation, they disagreed about the flow of the trends.
Our results indicate that although there is a great deal of disagreement among the participants regarding the underlying trends, the individual participants seemed to be fairly certain about the number of trends and their tracing.
Thus it is important to support users in identifying individual trends and avoiding wrong perceptions.
\section{Discussion and Conclusion}
In this paper, we presented a novel coloring method for line-based density plots. 
Although our approach is an image-based analysis of line data, it could be applied to density plots of parallel coordinates or even other types of data.
In contrast to the in Section~\ref{sec:related} introduced line-based methods, which group similar lines by colorization of the lines, abstract them with statistical information of ensembles, or combine them by altering their paths using edge-bundling, we aggregate the individual line into line-based density plots and enrich them by clustering similar regions in the image space.
 
\noindent \textbf{Future works and limitations.}
The basic unit of data for our presented method is pixels, which means the higher the resolution of the visualization, the more computationally demanding it is.
However, it is unnecessary to perform binning for each pixel. Different bin sizes can be used in different areas depending on the distribution of the data. For example, a larger bin size can be used for unimportant areas.
So the plot could be divided into differently-sized bins to reduce the number of bins.
Therefore, the next step would be to explore automatic methods for finding the best binning strategy.
In subsequent works, we also want to reduce the number of clustering errors by investigating other sampling strategies.

\revised{Edge bundling is a widely used approach to address the clutter in line-based visualizations. Recent methods address the issue of ambiguities introduced by edge bundling. Our method could also be applied to disambiguate further the results generated by edge bundling methods.
}

\revised{With an increasing number of clusters, the angle between assigned colors on the hue wheel decreases, which reduces their distinguishability. Already for 7 different colors (see Fig.~\ref{fig:Greece-colored}), the colors start to be barely distinguished.
Furthermore, perception depends on cluster sizes, spacing of hues, and positional relationships between clusters. Conducting a user study could shed light on these aspects.
}

\revised{Our small-scale motivational study was only concerned with showing whether and to what extent the perception of trends in line-based density plots is ambiguous. It encompassed only university students and Ph.D. candidates. But it underpinned our hypothesis that line-based density plots are indeed perceived ambiguously by users. We investigate only whether the user’s understanding of line-based density plots is ambiguous. Therefore, we did not include a comparison of line-based coloring methods and our method in the user study. A further study should have a larger and more diverse population, including perception and understanding of our results.}

Currently, we summarize three types of ambiguities in line-based density plots, but it is possible that additional types also exist. In future work, we intend to develop a more comprehensive taxonomy to categorize the ambiguities found in line-based density plots.

\section*{Figure Credits}
\label{sec:figure_credits}
For Fig.~\ref{fig:usecase_Greece}, the gray background image of Greece was taken from OpenStreetMaps~\cite{OpenStreetMap} at location \url{https://www.openstreetmap.org/#map=7/37.055/22.313&layers=T}.

\acknowledgments{%
	This work was supported in part by Deutsche Forschungsgemeinschaft (DFG) Project 410883423, Project 251654672 – TRR 161 ``Quantitative methods for visual computing'', KE 740/17-2 of the FOR2111 ``Questions at the Interface'', the National Key R\&D Program of China (2022ZD0160805), NSF China (No.62132017, 62141217), and the Shandong Provincial Natural Science Foundation (No. ZQ2022JQ32).%
}

\bibliographystyle{abbrv-doi-hyperref}

\bibliography{template}



\end{document}